\documentclass[11pt,a4paper]{article}
\pdfoutput=1

\usepackage[colorlinks=true, linkcolor=black!50!blue, urlcolor=blue, citecolor=blue, anchorcolor=blue]{hyperref}
\usepackage[font=small,labelfont=bf,margin=0mm,labelsep=period,tableposition=top]{caption}
\usepackage[a4paper,top=1.5cm,bottom=2cm,left=1.5cm,right=1.5cm,bindingoffset=0mm]{geometry}

\usepackage{tabularx}
\usepackage{placeins,cite}
\usepackage{graphicx}
\usepackage{soul}
\usepackage{subcaption}
\usepackage{float}
\usepackage{afterpage}
\usepackage{epsfig}
\usepackage{amssymb}
\usepackage{amsmath}
\usepackage{bm}
\usepackage{multirow}
\usepackage{url}
\usepackage{xcolor}
\usepackage{url}
\usepackage{booktabs,multirow}
\usepackage{textcomp}
\usepackage{comment}
\graphicspath{{../}{./figures/}}

\usepackage{makecell}

\makeatletter
\def\thickhline{%
             \noalign{\ifnum0 =`}\fi\hrule \@height \thickarrayrulewidth \futurelet
             \reserved@a\@xthickhline}
\def\@xthickhline{\ifx\reserved@a\thickhline
                \vskip\doublerulesep
                \vskip -\thickarrayrulewidth
                \fi
                \ifnum0 =`{\fi}}
\makeatother
\newlength{\thickarrayrulewidth}
\usepackage{tikz}
\usetikzlibrary{positioning}
\usetikzlibrary{shapes.geometric, arrows}
\usetikzlibrary{arrows.meta}
\usepackage{varwidth}
\usepackage{xcolor}
\definecolor{mtplotlib1}{HTML}{1f77b4}
\definecolor{mtplotlib2}{HTML}{ff7f0e}
\definecolor{mtplotlib3}{HTML}{2ca02c}
\definecolor{mtplotlib4}{HTML}{d62728}

\tikzset{%
  >={Latex[width=2mm,length=2mm]},
            base/.style = {rectangle, rounded corners, draw=black,
                           minimum width=4cm, minimum height=1cm,
                           text centered}, %, font=\sffamily},
            mystyle/.style={rectangle, rounded corners, draw=black,
            minimum width=12cm, minimum height=1cm,
            text centered}, %, font=\sffamily},
    col0/.style = {base, fill=white!30},
    col1/.style = {base, fill=mtplotlib1!30},
    col11/.style = {mystyle, fill=mtplotlib1!30},
    col2/.style = {base, fill=mtplotlib2!30},
    col3/.style = {base, fill=mtplotlib3!30},
    col4/.style = {base, minimum width=2.5cm, fill=mtplotlib4!15,}%font=\ttfamily},
}

\RequirePackage[normalem]{ulem} 

\newcommand{\be}{\begin{equation}}
\newcommand{\ee}{\end{equation}}
\newcommand{\bea}{\begin{eqnarray}}
\newcommand{\eea}{\end{eqnarray}}
\newcommand{\bi}{\begin{itemize}}
\newcommand{\ei}{\end{itemize}}
\newcommand{\ben}{\begin{enumerate}}
\newcommand{\een}{\end{enumerate}}
\newcommand{\la}{\left\langle}
\newcommand{\ra}{\right\rangle}
\newcommand{\lc}{\left[}
\newcommand{\rc}{\right]}
\newcommand{\lp}{\left(}
\newcommand{\rp}{\right)}

\def\frac#1#2{{{#1}\over {#2}}}
\def\gsim{\mathrel{\rlap{\lower4pt\hbox{\hskip1pt$\sim$}}
    \raise1pt\hbox{$>$}}}       
\def\lsim{\mathrel{\rlap{\lower4pt\hbox{\hskip1pt$\sim$}}
    \raise1pt\hbox{$<$}}}

\newcommand{\ifb}{\mathrm{fb}^{-1}}

\newcommand{\draft}[1]{}
\def\beq{\begin{equation}}
\def\eeq{\end{equation}}

\numberwithin{equation}{section}
\numberwithin{figure}{section}
\numberwithin{table}{section}

\usepackage{tabularx}
\newcolumntype{C}[1]{>{\centering\arraybackslash}p{#1}}

\definecolor{darkblue}{rgb}{0.0,0,0.5}
\definecolor{darkgreen}{rgb}{0.0,0.3,0.0}
\definecolor{redish}{rgb}{0.675,0,0.2}
\definecolor{red}{rgb}{0.8,0,0}
\definecolor{green}{rgb}{0,0.6,0}
\definecolor{bluish}{rgb}{0.2,0.2,0.675}
\definecolor{mygrey}{rgb}{0.6,0.6,0.6}

\usepackage{tikz} 
\usetikzlibrary{shapes,arrows,positioning,automata,backgrounds,calc,er,patterns,arrows.meta}
\usepackage{tikz-feynman}
\tikzfeynmanset{compat=1.0.0}
\usepackage{varwidth}
\usepackage{xcolor}
\definecolor{mtplotlib1}{HTML}{1f77b4}
\definecolor{mtplotlib2}{HTML}{ff7f0e}
\definecolor{mtplotlib3}{HTML}{2ca02c}
\definecolor{mtplotlib4}{HTML}{d62728}

\tikzset{%
  >={Latex[width=2mm,length=2mm]},
            base/.style = {rectangle, rounded corners, draw=black,
                           minimum width=4cm, minimum height=1cm,
                           text centered}, %, font=\sffamily},
            mystyle/.style={rectangle, rounded corners, draw=black,
            minimum width=12cm, minimum height=1cm,
            text centered}, %, font=\sffamily},
    col0/.style = {base, fill=white!30},
    col1/.style = {base, fill=mtplotlib1!30},
    col11/.style = {mystyle, fill=mtplotlib1!30},
    col2/.style = {base, fill=mtplotlib2!30},
    col3/.style = {base, fill=mtplotlib3!30},
    col4/.style = {base, minimum width=2.5cm, fill=mtplotlib4!15,}%font=\ttfamily},
}

\usepackage{tabularx}
\usepackage{subcaption}
\newcolumntype{C}[1]{>{\centering\arraybackslash}p{#1}}

\definecolor{lightblue}{rgb}{0.0,0.5,1.0}

\begin{document}
\newgeometry{top=1.5cm,bottom=1.5cm,left=1.5cm,right=1.5cm,bindingoffset=0mm}

\begin{center}
  {\Large \bf A  First Determination of the LHC Neutrino Fluxes from FASER Data }\\
  \vspace{1.1cm}
  {\small
  Jukka John$^1$, Felix Kling$^{2,3}$, Jelle Koorn$^1$, Peter Krack$^{1,4}$ 
  and Juan Rojo$^{1,4}$
  }\\
  
\vspace{0.7cm}

{\it \small
  ~$^1$Nikhef Theory Group, Science Park 105, 1098 XG Amsterdam, The Netherlands\\[0.1cm]
   ~$^2$Deutsches Elektronen-Synchrotron DESY, Notkestr. 85, 22607
Hamburg, Germany\\[0.1cm]
   ~$^3$Department of Physics and Astronomy, University of California, Irvine, CA 92697-4575, USA\\[0.1cm] 
    ~$^4$Department of Physics and Astronomy, Vrije Universiteit, NL-1081 HV Amsterdam, The Netherlands\\[0.1cm]  
 }

\vspace{1.0cm}

{\bf \large Abstract}

\end{center}

The detection of TeV neutrinos from the LHC by the far-forward detectors FASER and SND@LHC enables a plethora of novel physics opportunities.
Among these, the measurement of the flavour, energy, and rapidity dependence of the LHC forward neutrino fluxes provides unique constraints on theoretical predictions of forward particle production in hadronic collisions. 
We demonstrate that neutrino event yield measurements at FASER from Run 3 and at its HL-LHC upgrades enable a theory-agnostic extraction of the LHC forward neutrino fluxes.
We exploit the equivalence of the problem with the determination of parton distributions from deep-inelastic structure functions to apply the NNPDF approach, based on machine learning regression and the Monte Carlo replica method, to LHC neutrino fluxes.  
The resulting NN$\nu$flux methodology is validated through closure tests and applied to a first extraction of the LHC muon neutrino flux from the FASER 2024 data.
We show how NN$\nu$flux can discriminate between event generators of forward hadron production;  scrutinize a possible intrinsic charm component in the proton; and constrain BSM scenarios with enhanced decays of neutral hadrons into neutrinos.

\clearpage

\tableofcontents

\section{Introduction}

The recent observation of LHC neutrinos, first by FASER~\cite{FASER:2023zcr} and shortly thereafter by SND@LHC~\cite{SNDLHC:2023pun}, followed by subsequent measurements of inclusive and energy-differential cross-sections by FASER/FASER$\nu$~\cite{FASER:2024ref, FASER:2024hoe}, have kick-started the era of collider neutrino physics~\cite{Anchordoqui:2021ghd, Feng:2022inv, Adhikary:2024nlv, FPFWorkingGroups:2025rsc, Ariga:2025qup, Kling:2025zon}.
Measurements of neutrino scattering rates at these far-forward LHC detectors provide unique information on neutrino interactions at TeV energies, on the quark and gluon substructure of protons and nuclei, and on the mechanisms driving forward hadron production in proton-proton collisions.
Furthermore, neutrino-initiated deep-inelastic scattering (DIS) measurements can be complemented by the information provided by their muon-initiated counterparts~\cite{Francener:2025pnr} at the same far-forward detectors.

Assuming that neutrino-nucleon interactions can be evaluated in perturbative QCD, measurements of neutrino (and muon) scattering rates at the LHC far-forward detectors constrain light and heavy hadron production in high-energy proton-proton collisions~\cite{Kling:2023tgr}, opening novel windows to QCD dynamics and proton structure in kinematic regions which are mostly unexplored~\cite{Bhattacharya:2023zei}. 
A first study in this direction was carried out by FASER~\cite{FASER:2024ref}, where muon neutrino flux measurements from the electronic detector were used to validate the predictions of different Monte Carlo (MC) generators of forward hadron production.
In particular, the ratio between neutrinos from pion and from kaon decays, a key ingredient of possible solutions for the muon puzzle in cosmic rays~\cite{Anchordoqui:2022fpn, Sciutto:2023zuz}, was investigated.  

Building upon these initial studies, the aim of this work is to develop a systematic framework for a data-driven, theory-agnostic determination of the forward neutrino fluxes from event yield measurements at FASER and SND@LHC, their upgrades at the HL-LHC, and the detectors of the proposed Forward Physics Facility (FPF)~\cite{Anchordoqui:2021ghd, Feng:2022inv, Adhikary:2024nlv, FPFWorkingGroups:2025rsc}. 
We achieve this by exploiting the equivalence of the problem with the determination of the nucleon's parton distributions functions (PDFs)~\cite{Gao:2017yyd, Kovarik:2019xvh, Ethier:2020way, Amoroso:2022eow} from DIS structure functions.
Indeed, predictions for fiducial neutrino event rates at FASER can be expressed~\cite{Cruz-Martinez:2023sdv} as a convolution of the neutrino fluxes with a QCD kernel accounting for acceptance and selection cuts.
This structure is analogous to DIS structure functions, evaluated in terms of an interpolated perturbative QCD kernel~\cite{Bertone:2013vaa, Candido:2022tld, Salam:2008qg} convoluted with the non-perturbative PDFs.

This correspondence between the determination of the LHC neutrino fluxes and that of PDFs from DIS data suggests that methodologies used for the latter should also be applicable to the former.
Here we apply the NNPDF formalism~\cite{NNPDF:2014otw, NNPDF:2021njg, Candido:2023utz} to the determination of the LHC neutrino fluxes from current and future data taken at the LHC far-forward detectors.
Our approach, designated as NN$\nu$flux, entails combining neural networks for unbiased regression with the Monte Carlo replica method for faithful uncertainty estimate.
We focus on constraining the energy dependence of the fluxes, though our approach is general and can be extended to multi-differential parametrisations also accounting for their rapidity dependence~\cite{Kling:2021gos, FASER:2024ykc, FASER:CERN-FASER-CONF-2025-001}.

The NN$\nu$flux methodology is validated through closure tests~\cite{DelDebbio:2021whr, Ball:2025xgq} with synthetic data generated from a known ground truth for the neutrino fluxes. 
These closure tests are satisfied irrespective of the choice of the kinematic distribution used as input, the baseline flux calculation, the detector geometry, and the integrated luminosity. 
We apply NN$\nu$flux to the  measurements of muon neutrino event yields from the FASER 2024 dataset~\cite{FASER:2024ref} based on $\mathcal{L}_{\rm pp}=65.6$~fb$^{-1}$ to carry out a first determination of the muon neutrino flux differential in $E_\nu$ and its associated uncertainties. 
We also quantify the statistical pull between the FASER data and generators modelling forward hadron production, some of which are strongly disfavoured. 

The NN$\nu$flux approach streamlines the connection between calculations of forward neutrino production in hadronic collisions, on the one hand, and measurements at the LHC far-forward detectors, on the other hand. 
To highlight representative physics opportunities we apply NN$\nu$flux to discriminate between predictions from event generators for forward hadron production; to investigate a possible intrinsic charm component in the proton; and to constrain BSM scenarios with enhanced branching fractions of neutral hadrons into neutrinos.
In addition to the FASER measurements of~\cite{FASER:2024ref}, we perform NN$\nu$flux projections for electron and muon neutrino fluxes at FASER$\nu$ for $\mathcal{L}_{\rm pp}=150$~fb$^{-1}$ (Run 3) and 3~ab$^{-1}$ (HL$-$LHC) as well as for FASER$\nu$2 at the FPF (also 3~ab$^{-1}$).
Beyond these initial applications, our method to extract neutrino fluxes can validate predictions of forward $D$-meson production to constrain prompt neutrino flux predictions~\cite{Gauld:2015kvh, Bai:2022xad} at neutrino telescopes such as IceCube~\cite{IceCube:2014stg} and KM3NET~\cite{KM3Net:2016zxf}.

The structure of the paper is as follows.
In Sect.~\ref{sec:formalism} we present an overview of neutrino flux predictions, the fast interface to the {\sc\small POWHEG+Pythia8} DIS simulations, and the NN$\nu$flux fitting methodology together with its validation through closure tests. 
The NN$\nu$flux formalism is applied to the FASER 2024 measurements in~Sect.~\ref{sec:results_faser} to achieve a first data-driven extraction of the LHC muon neutrino fluxes.
Sect.~\ref{sec:interpretation} presents results for three representative QCD and BSM applications of our method.
In Sect.~\ref{sec:summary}
we summarise and present a possible outlook for future developments.
App.~\ref{app:2d-lumis} discusses how the FK-table interpolation formalism can be extended to the case of both energy- and rapidity-dependent neutrino fluxes. 

\section{The NN$\nu$flux formalism}
\label{sec:formalism}

Here we provide an overview of neutrino flux predictions, present the theoretical formalism underpinning neutrino scattering at the LHC far-forward neutrino experiments, describe how neutrino event yields can be evaluated in terms of a convolution between an interpolated kernel and the neutrino fluxes, and discuss how to extract the neutrino fluxes from the measured event yields by means of machine learning regression.

\subsection{Neutrino flux predictions}
\label{eq:overview_fluxes}

First we present an overview of the LHC neutrino flux predictions considered in this work.
These fluxes differ on the treatment of the QCD production mechanisms, the type and geometry of the considered detector, and the proton-proton luminosity at which this detector is exposed. 

The first reckoning that high-energy proton-proton collisions generate intense, strongly collimated, and highly energetic beams of neutrinos in the forward direction was in the early 1980s~\cite{DeRujula:1984ns} and conceptually discussed again several times in the following decades~\cite{Winter:1990ry, DeRujula:1992sn, Vannucci:1993ud, Park:2011gh, Beni:2019gxv, Foldenauer:2021gkm}. 
However, only in the 2020, the first experiments, FASER~\cite{FASER:2022hcn} and SND@LHC~\cite{SNDLHC:2022ihg}, were installed to exploit this opportunity. 
Both experiments are located about 480~m downstream at opposite sites of the ATLAS interaction point, inside previously unused service tunnels from the LEP era. 
These locations allow the experiments to be placed in or close to the center of the neutrino beam, thereby maximizing the event rates. 
Both FASER and SND@LHC have been taking data since the summer of 2022 and have since then reported the first observation of collider neutrinos~\cite{FASER:2023zcr, SNDLHC:2023pun}. 
In addition, FASER has further measured the neutrino flux for the first time~\cite{FASER:2024ref}. 

The FASER detector is aligned with the beam collision axis and covers pseudorapidities $\eta \gtrsim 8.5$. 
Located at its front is the FASER$\nu$ neutrino detector, which consists of a tungsten target with roughly 1.2 tons target mass that is interleaved with emulsion films~\cite{FASER:2019dxq, FASER:2020gpr}. 
This detector provides a high resolution image of the charged particle tracks produced in neutrino interaction and allows the identification of the neutrino flavor as well as the measurement of their energy~\cite{FASER:2021mtu, FASER:2025qaf}.
A first observation of neutrinos with FASER$\nu$ has been reported in~\cite{FASER:2024hoe}. 
Located before FASER$\nu$ is a veto system to identify the presence of incoming particles. 
Located behind are FASER's spectrometer and calorimeter system~\cite{FASER:2018ceo, FASER:2018bac}. 
While the electronic detector is optimized for searches for long-lived particle decays~\cite{Feng:2017uoz}, it also acts as a spectrometer for muons produced in neutrino interactions. 
The FASER collaboration has utilized this electronic detector components to observe muon neutrinos via a muon appearance signature and measure of the muon neutrino and anti-neutrino flux as a function of the neutrino energy~\cite{FASER:2023zcr, FASER:2024ref}. 
The SND@LHC detector is located on the other side of ATLAS at an off-axis location covering the pseudorapidity range $7.2 < \eta < 8.4$.
Due to its off-axis location and smaller overlap with the neutrino beam, it collects a roughly ten times smaller number of neutrino events~\cite{Ariga:2025qup}. 

Both FASER and SND@LHC will continue to operate until the end of Run~3 of the LHC in 2026. 
Upgrades are envisioned for operation in Run~4 starting in 2030 with similarly sized detectors at the same locations~\cite{FASER:2025myb, Abbaneo:2926288}. 
In addition, a continuation and expansion of this experimental program during the HL-LHC era with significantly larger detectors has been proposed in the context of the FPF~\cite{Anchordoqui:2021ghd, Feng:2022inv, Adhikary:2024nlv}.
The FPF proposal consist of a dedicated cavern to be constructed 620 m downstream of ATLAS and designed to accommodate a suite of experiments. 
In addition to FASER2 for long-lived particle searches and FORMOSA for millicharged particles searches~\cite{Foroughi-Abari:2020qar}, the FPF encompasses two dedicated neutrino experiments, FLArE and FASER$\nu$2, with FLArE using a liquid argon time projection chamber with 10~ton target mass followed by a calorimeter and FASER$\nu$2 using a 20~ton tungsten target interleaved with emulsion films. 
Located in the back of the FPF is FASER2, which acts as muon spectrometer for the two neutrino experiments. 

In this work we consider four different data-taking and detector settings to analyse the potential of the LHC far-forward experiments to constrain the neutrino fluxes: 
\begin{description}
\item[FASER (2024).] The first measurement of the muon neutrino flux by FASER was presented in 2024~\cite{FASER:2024ref}. 
The analysis is based on data collected during 2022 and 2023 and corresponding to an integrated luminosity of $\mathcal{L}_{\rm pp}=65.6$~fb$^{-1}$.
It uses a cylindrical fiducial volume with a radius of 10~cm and an equivalent depth of 88.5~cm of tungsten, matching the transverse dimensions of the spectrometer and positioned directly in front of it, yielding an effective target mass of 537~kg. 

\item[FASER$\nu$ at Run~3.] 
The main neutrino analysis at FASER will be performed using the FASER$\nu$ emulsion detector. 
FASER$\nu$ has dimensions of $25~\text{cm} \times 30~\text{cm} \times 80~\text{cm}$ of tungsten, providing a target mass of 1.1~tons. 
By the end of Run~3, the detector is expected to collect about $\mathcal{L}_{\rm pp}=150~\ifb$ of data which provides a sample of around $10^4$ neutrino interactions~\cite{FASER:2024ykc}. 

\item[FASER$\nu$ at the HL-LHC.] 
The FASER experiment has been approved to continue its operation in LHC Run~4~\cite{Boyd:2882503} and plans to operate with an upgraded neutrino detector with similar location and dimensions to the current FASER$\nu$ setup~\cite{FASER:2025myb}. 
To estimate the sensitivity of this continued operation, we consider a detector with the same dimensions and location as FASER$\nu$ exposed to an integrated luminosity of $\mathcal{L}_{\rm pp}=3~\text{ab}^{-1}$. 

\item[FASER$\nu$2 at the FPF.] 
To illustrate the sensitivity of substantially larger neutrino detectors at the FPF, we consider FASER$\nu$2. 
The detector would be located 620~m downstream of ATLAS and has dimensions of $40~\text{cm} \times 40~\text{cm} \times 6.8~\text{m}$ of tungsten, aligned with the center of the neutrino beam and providing a target mass of 20~tons. 
Also here we assume an integrated luminosity of $3~\text{ab}^{-1}$. 
\end{description}

\noindent
The forward neutrino beam at the LHC and reaching the far-forward detectors primarily originates from the decays of pions, kaons and charm hadrons, with a small contribution from beauty hadrons.
To simulate this flux, we follow the method outlined in~\cite{Kling:2021gos, FASER:2024ykc}. 
To simulate the production of light hadrons, we use the hadronic interaction model {\sc\small EPOS-LHC}~\cite{Pierog:2013ria}. 
To model the production of charm hadrons we use {\sc\small POWHEG}~\cite{Nason:2004rx, Frixione:2007vw, Alioli:2010xd} matched with {\sc\small PYTHIA~8.3}~\cite{Bierlich:2022pfr} for parton shower and hadronization as described in~\cite{Buonocore:2023kna}
and based on NNPDF3.1sx+LHCb~\cite{Ball:2017otu, Bertone:2018dse}.
We also show results for forward light hadron production modelled using {\sc\small SIBYLL~2.3d}~\cite{Riehn:2019jet}, {\sc\small QGSJET~2.04}~\cite{Ostapchenko:2010vb} and {\sc\small DPMJET~2019}~\cite{Fedynitch:2015kcn} as well as charm production obtained using {\sc\small SIBYLL~2.3d} and {\sc\small DPMJET~2019}.
The fast neutrino flux simulation introduced in~\cite{Kling:2021gos} with the LHC Run~3 configuration as described in~\cite{FASER:2024ykc} was used to model the propagation of light long-lived mesons though the LHC's beam pipe and magnetic fields as well as to simulate their decays. 
Additional flux calculations, tailored to specific physical applications, are introduced in Sect.~\ref{sec:interpretation}.

When generating synthetic data, we will consider only statistical uncertainties for the binned neutrino scattering event yields, and no attempt is made to model systematic uncertainties.
The exception is the fit to synthetic data based on the FASER 2024 measurements binned in $E_\nu$, for which the full covariance matrix of the measurement is available~\cite{FASER:2024ref} and hence used.
We emphasize that the application of NN$\nu$flux to experimental data, as done in Sect.~\ref{sec:results_faser}, should always include all relevant sources experimental and theoretical uncertainties as well as their correlations.
 
\subsection{Neutrino DIS at the LHC}
\label{subsec:nuDIS}

In this work we consider (anti-)neutrino DIS off a fixed target,
\be
\nu_i~(\bar{\nu}_i) + N \to \ell^-(\ell^+) + X_h \, ,\quad i = e,\mu, \tau \, ,
\ee
where $N$ indicates the target nucleus and $X_h$ the hadronic final state. 
We assume a measurement of
the final-state charged lepton energy $E_\ell$
and scattering angle $\theta_\ell$ and of the energy of the hadronic
final state $E_h$.
The measurement of these three final-state
variables enables\footnote{The kinematic mapping in Eq.~(\ref{eq:dis_kinematic_mapping}) assumes neutrinos with $p_{\nu,\rm T}=0$.
By measuring the position of the neutrino introduction vertex, the neutrino transverse momentum can be inferred and corrected. 
Corrections due to $p_{\nu,\rm T}\ne 0$ are nevertheless small e.g. for the FASER$\nu$ kinematics one has that $p_{\nu,\rm T}\lsim $ 0.8 GeV.} the reconstruction
of the neutrino energy $E_\nu$
and of the DIS variables $Q^2$ and $x$:
\bea
 E_\nu &=& E_h + E_\ell \, , \nonumber \\
 Q^2 &=& 4 ( E_h + E_\ell) E_\ell \sin^2 \lp \theta_\ell/2\rp \, ,  \label{eq:dis_kinematic_mapping}\\
 x&=& 4 ( E_h + E_\ell) E_\ell \sin^2 \lp \theta_\ell/2\rp/2m_N E_h \, ,\nonumber
 \eea
 with $m_N$ being the nucleon mass.
 
 In addition to the energy $E_\nu$, the appraisal of QCD production mechanisms for forward neutrinos benefits from accessing the dependence on their rapidity  $y_\nu$, given by
\be
y_\nu = \frac{1}{2}\ln \lp \frac{E_\nu + p_{\nu,z}}{E_\nu - p_{\nu,z}}\rp \, ,
\ee
where the neutrino four-momentum is defined
as 
\be
\label{eq:nu_mom_param}
p_\nu^\mu = \lp E_\nu, \vec{p}_{\nu,{\rm T}} ,p_{\nu,z}\rp \, ,
\ee
separating the $z$-component, aligned with the line of sight (LoS), from the transverse component, perpendicular to the beam direction.
For massless particles,
the rapidity coincides with the pseudo-rapidity
\be
y_\nu = \eta_\nu = -\ln \lp \tan \frac{\theta_\nu}{2}\rp \, ,
\ee
with $\theta_\nu$ being the angle of the neutrino momentum relative to the beam axis.
The coverage of a given detector in the $y_\nu$ depends on its distance from the primary proton-proton collision and of its positioning with respect to the LoS.
For instance  FASER$\nu$, placed at $z=480$~m downstream of the ATLAS interaction point and covering the transverse region up to $r\sim 20$~cm of the LoS, can collect neutrinos with rapidities $\eta_{\nu}\gsim \eta_{\nu}^{\rm (min)}\sim 8.5$,
which correspond to neutrino angles $\theta_\nu \lsim 4\times 10^{-4}$.  

The neutrino fluxes reaching the LHC far-forward detectors can be binned  in a double differential distribution in $(E_\nu,y_\nu)$, 
\be
\label{eq:2dfluxes_def}
\frac{d^2 N_{\nu_i}(E_\nu,y_\nu)}{dE_\nu dy_{\nu} }  \, ,\quad i=e,\mu,\tau \, ,
\ee
and determine how many neutrinos
pass through the detector transverse cross-section.
When integrating over neutrino rapidities we have 
\be
\label{eq:1dfluxes_def}
\frac{d N_{\nu_i}(E_\nu)}{dE_\nu }= \int_{y_\nu^{\rm (min)}}^\infty dy_\nu \frac{d^2 N_{\nu_i}(E_\nu,y_\nu)}{dE_\nu dy_{\nu} } f_{\phi}(y_\nu)\, ,
\ee
assuming a detector centred at the LoS.
In Eq.~(\ref{eq:1dfluxes_def}), and where $f_{\phi}(y_\nu)$ is a factor accounting for the finite coverage in the azimuthal angle $\phi$ as a function of $\eta_\nu$.
For an off-axis detector such as SND@LHC one has 
\be
\frac{d N_{\nu_i}(E_\nu)}{dE_\nu }= \int_{y_\nu^{\rm (min)}}^{y_\nu^{\rm (max)}} dy_\nu \frac{d^2 N_{\nu_i}(E_\nu,y_\nu)}{dE_\nu dy_{\nu} } f_{\phi}(y_\nu) \, ,
\ee
with the coverage in rapidity being determined
by the detector geometry.
In the following, we focus on on-axis spherically symmetric detectors and set $f_\phi(y_\nu)=1$.

\paragraph{Even rate calculation.}
The most general observable that one can consider in (inclusive) neutrino DIS at the LHC is the number of  charged-current interactions for a given binning in
\be
\lp E_h, E_\ell, \theta_\ell, y_\nu\rp \quad {\rm or~alternatively~in} \quad \lp x,Q^2, E_\nu, y_\nu\rp \, .
\ee
Hence we want to predict
\be
\nonumber
 N_{\rm int}^{(\nu_i)}(x,Q^2,E_\nu, y_\nu)
 \equiv N_{\rm int}^{(\nu_i)}\lp x_{\rm min}\le x \le x_{\rm max}, Q^2_{\rm min}\le Q^2 \le Q^2_{\rm max}, E_\nu^{\rm min}\le E_\nu \le E_\nu^{\rm max}, y_\nu^{\rm min}\le y_\nu \le y_\nu^{\rm max}\rp 
\ee
within fiducial acceptance.
Reconstructing the kinematic information from these different variables provides complementary information.
Indeed, measurement of the $(E_\nu,y_\nu)$ dependence  provides direct information on the properties of the incoming neutrino fluxes, while the measurements of
the $(x,Q^2)$ dependence mostly constrain the neutrino scattering cross-section.
The event rate can also
be expressed  as
\be
\nonumber
 N_{\rm int}^{(\nu_i)}(E_\ell,\theta_\ell, E_h, y_\nu)
 \equiv N_{\rm int}^{(\nu_i)}\lp E_\ell^{\rm min}\le E_\ell \le E_\ell^{\rm max}, \theta_\ell^{\rm min}\le \theta_\ell \le \theta_\ell^{\rm max}, E_h^{\rm min}\le E_h \le E_h^{\rm max}, y_\nu^{\rm min}\le y_\nu \le y_\nu^{\rm max}\rp \, ,
\ee
in terms of final-state variables.
The optimal binning should be determined on a case by case basis depending on the available statistics and the desired physics application.\footnote{One may also consider unbinned measurements which retain the full event-by-event information, such as those based on neural simulation-based inference methods applied by ATLAS in~\cite{ATLAS:2025clx} and related approaches. }

The integrated event yields for a given bin will thus be given by 
\be
  \label{eq:event_yields_calculation}
 N_{\rm int}^{(\nu_i)}(x,Q^2,E_\nu, y_\nu)=\int_{Q^{2}_{\rm min}}^{Q^{2}_{\rm max}}
  dQ^2
 \int_{x_{\rm min}}^{x_{\rm max}}
  dx 
 \int_{E_\nu^{\rm (min)}}^{E_\nu^{\rm (max)}}
 dE_{\nu}
 \int_{y_\nu^{\rm (min)}}^{y_\nu^{\rm (max)}}
 dy_\nu\, \widetilde{N}_{\rm int}^{(\nu_i)}(x,Q^2,E_\nu, y_\nu)\, , 
\ee
with the integrand defined as 
\be
\label{eq:nint_integrand}
\widetilde{N}_{\rm int}^{(\nu_i)}(x,Q^2,E_\nu, y_\nu) \equiv
 n_T L_T  \frac{d^2N_{\nu_i}(E_\nu,y_\nu)}{dE_{\nu}dy_\nu}  \frac{d^2\sigma^{\nu_i A}(x,Q^2,E_{\nu})}{dxdQ^2}  {\cal A}(E_\ell,\theta_\ell,E_h) \, , 
\ee
where $n_T$ is the atomic density of the target
material, $L_T$ is its length, and the three other contributions are the double-differential neutrino flux Eq.~(\ref{eq:2dfluxes_def}), the double-differential DIS cross-section
which contains information on the neutrino-target interactions and that can be evaluated in perturbative QCD, and an acceptance function ${\cal A}(E_\ell,\theta_\ell,E_h)$ which restrict the events to the fiducial region of the detector. 
While Eq.~(\ref{eq:nint_integrand}) is more amenable to analytical calculations, in practice we use a Monte Carlo event generator to evaluate the final-state acceptance and hence we have 
\be
\label{eq:nint_integrand_v2}
\widetilde{N}_{\rm int}^{(\nu_i)}(x,Q^2,E_\nu, y_\nu) \equiv
 n_T L_T \frac{d^2N_{\nu_i}(E_\nu,y_\nu)}{dE_{\nu}dy_\nu} \frac{d^2\sigma_{\rm MC}^{\nu_i A}(x,Q^2,E_{\nu})}{dxdQ^2}\Bigg|_{\rm fid}\, , 
\ee
where the last term corresponds to fiducial cross-section computed with a Monte Carlo event generator, e.g. in this work our DIS simulations are based on {\sc\small POWHEG+Pythia8} to achieve NLO+LL accuracy.

Since  we are interested in extracting the neutrino fluxes, we assume that the DIS cross-section is known and integrate over the DIS variables 
\be
\nonumber
 N_{\rm int}^{(\nu_i)}(E_\nu, y_\nu)
 \equiv N_{\rm int}^{(\nu_i)}\lp  E_\nu^{\rm (min)}\le E_\nu \le E_\nu^{\rm (max)}, y_\nu^{\rm (min)}\le y_\nu \le y_\nu^{\rm (max)}\rp \, ,
\ee
which can be computed as
\be
\label{eq:event_yields_calculation_v3}
 N_{\rm int}^{(\nu_i)}(E_\nu, y_\nu)=
  \int_{E_\nu^{\rm (min)}}^{E_\nu^{\rm (max)}}
 dE_{\nu}
 \int_{y_\nu^{\rm (min)}}^{y_\nu^{\rm (max)}}
 dy_\nu\,
 \int_{Q^{2}_0}^{2m_NE_\nu}
  dQ^2
 \int_{Q^2/2m_NE_\nu}^{1}  
  dx \
 \widetilde{N}_{\rm int}^{(\nu_i)}(x,Q^2,E_\nu, y_\nu)\, , 
\ee
where the integration limits in $Q^2$ and $x$ are given by 
\be
(2~{\rm GeV})^2\le Q^2 \le 2m_NE_\nu \,,\qquad
x_0 (= Q^2/2m_NE_\nu) \le x \le 1 \, ,
\ee
with the contribution from events with $Q< Q_{\rm min}=2~{\rm GeV}$ being negligible for FASER energies~\cite{Candido:2023utz}.
An analogous expression holds when integrating over neutrino rapidity
\be
\label{eq:event_yields_enu}
 N_{\rm int}^{(\nu_i)}(E_\nu)
 \equiv N_{\rm int}^{(\nu_i)}\lp  E_\nu^{\rm (min)}\le E_\nu \le E_\nu^{\rm (max)}\rp \, ,
\ee
where
\be
\label{eq:event_yields_calculation_v4}
 N_{\rm int}^{(\nu_i)}(E_\nu)=
  \int_{E_\nu^{\rm (min)}}^{E_\nu^{\rm (max)}}
 dE_{\nu}
 \int_{y_\nu^{\rm (min)}}^{\infty}
 dy_\nu\,
 \int_{Q^{2}_{0}}^{2m_NE_\nu}
  dQ^2
 \int_{Q^2/2m_NE_\nu}^{1}
  dx 
 \widetilde{N}_{\rm int}^{(\nu_i)}(x,Q^2,E_\nu, y_\nu)\, .
\ee
The same treatment holds for binned event yields in terms of other variables, e.g. $E_h$, $E_\ell$, or $\theta_\ell$.

\paragraph{Simulation settings.}
The NLO+LL fiducial cross-sections entering Eq.~(\ref{eq:nint_integrand_v2}) are evaluated using simulations based on {\sc\small POWHEG} matched to {\sc\small Pythia8}.
In addition to the DIS cut of $Q_{\mathrm{min}}  = 2\;\mathrm{GeV}$  to ensure the validity of perturbative QCD, we impose the following final-state cuts: $E_\ell > 100 \;\mathrm{GeV}$, $n_{\mathrm{tr}} \ge 5$, and $\Delta \phi > \pi/2$. 
Here $n_{\rm tr}$ stands for number of charged tracks, namely the number of charged particles with an energy of at least $E_{\rm tr}=2\;\mathrm{GeV}$, 
and $\Delta \phi$ is the angle between the final-state lepton and the hadronic system defined as the sum of all charged tracks in the event.
No cut is imposed on the invariant mass of the hadronic system $W$ (which is highly correlated to $n_{\rm tr}$).
The same values of the input physical parameters as in~\cite{vanBeekveld:2024ziz} are used.
The \textsc{Pythia8} dipole shower is used for the parton shower, with hadronization modelled by the Monash 2013 tune~\cite{Skands:2014pea}. 
The quark and gluon content of the tungsten target is modelled using the PDF4LHC21 NNLO PDFs~\cite{PDF4LHCWorkingGroup:2022cjn}, imposing isospin symmetry and neglecting nuclear effects.

\subsection{Fast interpolation grids}
\label{subsec:fk_tables}

We aim to perform a determination of the neutrino fluxes, Eqns.~(\ref{eq:2dfluxes_def}) or~(\ref{eq:1dfluxes_def}), from the  measurements of neutrino binned event rates. 
The direct evaluation of event yields such as Eq.~(\ref{eq:event_yields_enu}) in the context of a regression problem is however too computational expensive, since theoretical predictions need to be recomputed from scratch for each iteration of the fit. 
To bypass this limitation, we encapsulate the fiducial neutrino scattering cross-section of Eq.~(\ref{eq:nint_integrand_v2}) into a precomputed interpolation table, such that event yields can be efficiently evaluated in terms of a matrix multiplication with the neutrino fluxes. 
Fast grid interpolation methods have extensively applied to precompute DGLAP evolution kernels~\cite{Bertone:2013vaa,Candido:2022tld,Salam:2008qg}, NLO and NNLO partonic cross-sections~\cite{Carrazza:2020gss,Bertone:2014zva,Carli:2010rw,Kluge:2006xs,Czakon:2017dip,yadism,Cruz-Martinez:2025ffa}, and their combination~\cite{Barontini:2023vmr}.
Here we apply the same techniques to the {\sc\small POWHEG+Pythia8} DIS calculation, enabling the implementation of the data from the LHC far-forward detectors into a PDF-like regression problem by means of the so-called FK-tables~\cite{Ball:2010de}.

Following~\cite{vanBeekveld:2024ziz}, we define a normalised neutrino flux in terms a   ``neutrino PDF'' given by 
\be
\label{eq:neutrino_pdf_definition}
f_{\nu_i}(x_{\nu})\equiv \frac{\sqrt{s_{\rm pp}}}{2} \frac{dN_{\nu_i}(E_\nu)}{dE_{\nu}} \, ,\qquad i = e,\mu,\tau \, .
\ee
where in analogy with the quark and gluon PDFs, we define a ``neutrino momentum fraction'' $x_{\nu}$ as
\be
x_{\nu} \equiv \frac{2E_\nu}{\sqrt{s_{\rm pp}}} \, ,\qquad 0 \le x_{\nu} \le 1 \, ,
\label{eq:xnu}
\ee
such that $x_\nu=1$ corresponds to the maximal energy that the neutrino may have when produced in a symmetric proton-proton collision at $\sqrt{s_{\rm pp}}$.
We perform here the derivation for the $E_\nu$-dependent neutrino fluxes Eq.~(\ref{eq:1dfluxes_def}), with the case in which also $y_\nu$ is accounted for described in App.~\ref{app:2d-lumis}.

Expressed in terms of this neutrino
PDF, the neutrino interaction event rates binned in the neutrino energy are given by 
\be
\label{eq:nint_integrand_v6}
\widetilde{N}_{\rm int}^{(\nu_i)}(x,Q^2,E_\nu) \equiv
 \frac{2n_T L_T}{\sqrt{s_{\rm pp}}}f_{\nu_i}(x_\nu)  \frac{d^2\sigma_{\rm MC}^{\nu_i A}(x,Q^2,E_{\nu})}{dxdQ^2}\Bigg|_{\rm fid}\, . 
\ee
We now introduce a suitable interpolation basis $\{I_\alpha(x)\}$ to express the neutrino PDF as 
\begin{equation}
\label{eq:interpolated_expression_fluxes}
    f_{\nu_i}(x_\nu) \simeq  \sum_{\alpha=1}^{n_{x}} f_{\nu_i}(x_{\nu,\alpha})I_\alpha(x_\nu) \, ,
\end{equation}
with $n_x$ the number of nodes in the $x_\nu$ interpolation grid.
Using Eq.~(\ref{eq:interpolated_expression_fluxes}) we can express Eq.~(\ref{eq:nint_integrand_v6}) as 
\be
\label{eq:nint_integrand_v8}
\widetilde{N}_{\rm int}^{(\nu_i)}(x,Q^2,E_\nu) = 
\frac{2 n_T L_T}{\sqrt{s_{\rm pp}}}   \sum_{\alpha=1}^{n_{x}} f_{\nu_i}(x_{\nu,\alpha}) 
I_\alpha(x_\nu)  \frac{d^2\sigma_{\rm MC}^{\nu_i A}(x,Q^2,E_{\nu})}{dxdQ^2}\Bigg|_{\rm fid} \, ,
\ee
where $\{f_{\nu_i}(x_{\nu,\alpha})\}$ indicates the array of neutrino PDF values sampled at the grid nodes in $x_\nu$. 
We can now write the integrated event yields as 
\be
 N_{\rm int}^{(\nu_i)}(E_\nu)= 
 \frac{2 n_T L_T}{\sqrt{s_{\rm pp}}} \sum_{\alpha=1}^{n_{x}} f_{\nu_i}(x_{\nu,\alpha}) \lp \int_{E_\nu^{\rm (min)}}^{E_\nu^{\rm (\rm max)}}
 dE_{\nu}
\int_{Q^{2}_{0}}^{Q^{2}_{\rm max}}
  dQ^2
 \int_{x_{\rm min}}^{1}
  dx  \, \lc 
I_\alpha(x_\nu) 
 \frac{d^2\sigma_{\rm MC}^{\nu_i A}(x,Q^2,E_{\nu})}{dxdQ^2}\Bigg|_{\rm fid} \rc \rp  \, ,
 \nonumber
\ee
which has the same structure as Eq.~(\ref{eq:event_yields_calculation_v4}) with the difference being the replacement of the  neutrino flux with the interpolation functions. 
For $n_{E_\nu}$ bins with limits $(E_{\nu,j}^{\rm (min)},E_{\nu,j}^{\rm (max)})$ one has
\be
\label{nint_after_fktable}
 N_{\rm int}^{(\nu_i)}(E_{\nu,j})=  
 \sum_{\alpha=1}^{n_{x}} f_{\nu_i}(x_{\nu,\alpha}) {\rm FK}_{\alpha,j} \, ,\qquad
 j=1,\ldots, n_{E_\nu} \, ,
\ee
in terms of a precomputed fast interpolation grid (FK-table) given by
\be
\label{eq:FKtable_def}
{\rm FK}_{\alpha,j} \equiv \frac{2n_T L_T}{\sqrt{s_{\rm pp}}}
\int_{E_{\nu,j}^{\rm (min)}}^{E_{\nu,j}^{\rm (max)}}
 dE_{\nu}
 \int_{Q_0^2}^{2m_NE_\nu}
  dQ^2
 \int_{Q^2/2m_NE_\nu}^{1}
  dx  \, \lc  
I_\alpha(x_\nu) 
 \frac{d^2\sigma_{\rm MC}^{\nu_i A}(x,Q^2,E_{\nu})}{dxdQ^2}\Bigg|_{\rm fid} \rc \, .
\ee
In this manner, for a given neutrino flux, the associated event yields can be efficiently evaluated from Eq.~(\ref{nint_after_fktable}), requiring only the values of the flux at the  grid nodes $\{ x_{\nu,\alpha}\}$ followed by a matrix multiplication with the  FK-table Eq.~(\ref{eq:FKtable_def}). 
The most computationally expensive part of the calculation, namely the NLO+LL simulations of the scattering cross-section in the fiducial detector region, is now precomputed and hence a single initial evaluation is necessary to use Eq.~(\ref{nint_after_fktable}) as input to a regression problem.
Similar expressions can be derived for other variables, such as the $E_\ell$, $E_h$, or $\theta_\ell$ distributions, in each case with their characteristic FK-table, as well as for multi-differential distributions.

\begin{figure}[t]
    \centering
\includegraphics[width=0.49\linewidth]{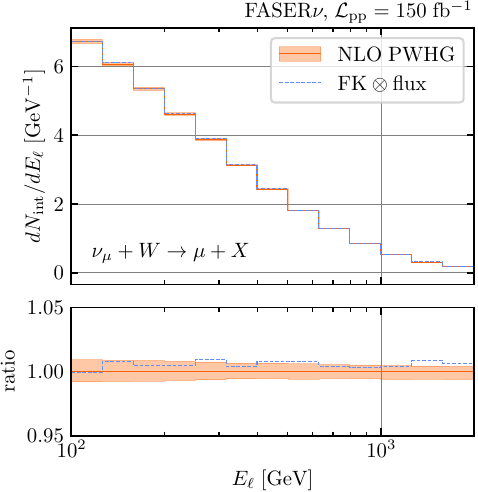}
\includegraphics[width=0.49\linewidth]{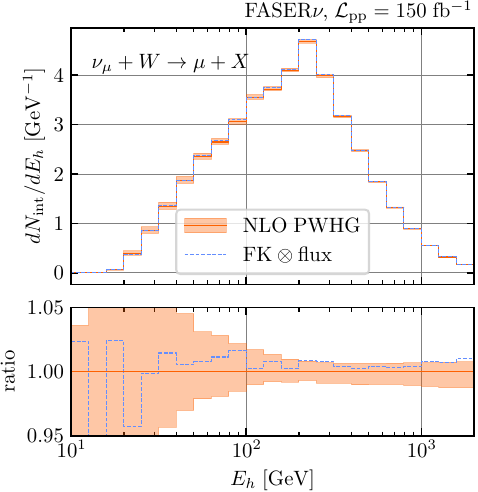}
    \caption{Comparison between the {\sc\small POWHEG} NLO predictions for the $E_\ell$ (left) and $E_h$ (right panel) distributions and their interpolated counterparts based on the FK-table formalism, Eq.~(\ref{nint_after_fktable}), for muon neutrino scattering at FASER$\nu$ for $\mathcal{L}_{\rm pp}=150$ fb$^{-1}$. 
    A common input neutrino flux is adopted in both cases.
    The bottom panels display the ratio with respect to the {\sc\small POWHEG} calculation, showing how the interpolation accuracy of the FK-tables reaches the sub-percent level in the kinematical region of phenomenological relevance.
    The uncertainty band stands for the scale variation of the {\sc\small POWHEG} calculation and is evaluated using the 7-point prescription for scale variations~\cite{vanBeekveld:2024ziz}.
     }
    \label{fig:fktable-validation}
\end{figure}

To validate the accuracy of the FK-table interpolation, Fig.~\ref{fig:fktable-validation} displays a comparison between the {\sc\small POWHEG} NLO predictions~\cite{Banfi:2023mhz,FerrarioRavasio:2024kem,vanBeekveld:2024ziz} for the $E_\ell$  and $E_h$  distributions and their interpolated counterparts based on the FK-table formalism at FASER$\nu$ for $\mathcal{L}_{\rm pp}=150$ fb$^{-1}$ for  muon neutrino scattering.
A common input neutrino flux is adopted in both cases.
The bottom panels display the ratio with respect to the {\sc\small POWHEG} calculation, showing how the interpolation accuracy of the FK-tables reaches the sub-percent level in the kinematical region of phenomenological relevance. 
The bands in the {\sc\small POWHEG+Pythia8} prediction estimate missing higher orders by means of the customary 7-point scale variation.
A similar level of agreement is found for other kinematic distributions and input neutrino fluxes considered in this analysis.

We take as basis of interpolation functions the logarithmic Lagrange interpolation polynomials, though our method also applies for other choices of interpolation bases. 
To achieve the sub-percent accuracy for the neutrino flux interpolation shown in Fig.~\ref{fig:fktable-validation}, following the approach used in the {\sc\small EKO} DGLAP evolution package~\cite{Candido:2022tld}, the grid in the neutrino momentum fraction $x_\nu$ is divided in blocks $B_\alpha=[ \alpha - \lfloor n/2 \rfloor , \alpha + \lfloor (n-1)/2 \rfloor]$
of a certain size $n$.
This choice is numerically advantageous since it implies  that the  interpolation polynomial is non-zero only for arguments in a limited range in $x_\nu$ in the neighbourhood of the node $\alpha$ and leads to smaller interpolation errors for the problem at hand. 

%%%%%%%%%%%%%%%%%%%%%%
%%%%%%%%%%%%%%%%%%

\subsection{Fitting methodology}
\label{sec:nu_flux_fitting}

Having established with Eq.~(\ref{nint_after_fktable}) that neutrino event yields, for a generic choice of kinematic variables, can be expressed in terms of a matrix multiplication between the neutrino flux and a pre-computed FK-table
\be 
N_{\rm int}^{(\nu_i)}(E_{\nu,j})= \sum_{\alpha=1}^{n_{x}}
 f_{\nu_i}(x_{\nu,\alpha}){\rm FK}_{\alpha,j} \,, \quad j=1,\ldots,n_{E_\nu}\,, 
\ee
we can exploit that this expression is analogous to the case of DIS structure functions~\cite{DelDebbio:2007ee,Ball:2010de,Ball:2008by},
\be 
F(x_i, Q^2_j)= \sum_{k=1}^{n_f}\sum_{\alpha=1}^{n_{x}}
 f_{k}(x_{\alpha},Q_0){\rm FK}_{\alpha k,ij} \,, \quad i=1,\ldots,n_x\,, 
 \quad j=1,\ldots,n_{Q^2}\,,
\ee
to deploy the NNPDF fitting methodology to the extraction of the LHC forward neutrino fluxes. 

To this end, we parametrise the neutrino PDFs as follows:
\be
\label{eq:NN_parametrisation}
f_{\nu_i}(x_{\nu}) = x_{\nu}^{a_i} (1-x_\nu)^{b_i}{\rm NN}_{\nu_i}(x_\nu) \, ,
\ee
with ${\rm NN}_{\nu_i}(x_\nu)$ being a feed-forward neural network and $a_i,b_i$ are preprocessing exponents introduced to speed up the training~\cite{Ball:2016spl} and  fitted alongside with the neural network parameters. 
The $b_i\ge 0$ exponent ensures the correct asymptotic behaviour of the parametrised fluxes, 
\be
 f_{\nu_i}(x_{\nu}\to 1) \to 0 \, ,
\ee
in the large-$x_\nu$ extrapolation region where there are no experimental constraints. 
The choice of neural networks is motivated by their property of operating as universal unbiased interpolants, which ensures they can reproduce any functional behaviour for the neutrino fluxes preferred by the data without the need of model assumptions. 
Here we provide results based on an independent parametrisation of the four fluxes
\be
f_{\nu_e}(x_{\nu})\,, f_{\bar{\nu}_e}(x_{\nu})\,, f_{\nu_\mu}(x_{\nu})\,, f_{\bar{\nu}_\mu}(x_{\nu})\, ,
\ee
and for detectors not sensitive to the lepton charge we sum over neutrino and antineutrino fluxes. 
Though not considered here, our methodology is also amenable to tau neutrino fluxes.

The uncertainties associated to the determination of the neutrino fluxes~Eq.~(\ref{eq:NN_parametrisation}) are estimated by means of the Monte Carlo replica method. 
Given $n_{\rm dat}$ experimental measurements $D_i$, one generates $N_{\rm rep}$ replicas $D_i^{(k)}$  by sampling the original data from a multi-Gaussian distribution such that
\be
\lim_{N_{\rm rep}\to\infty} \mathrm{cov}\left(D_i^{(k)}, D_j^{(k)}\right)={\rm cov}_{ij} \, ,\quad i,j=1,\ldots, n_{\rm dat} \, ,
  \label{covmat}
\ee
reproducing the covariance matrix of the data, which includes all relevant experimental and theoretical errors and their correlations.
One then fits Eq.~(\ref{eq:NN_parametrisation}) to each of the $N_{\rm rep}$ replicas, and the resulting sample $f_{\nu_i}^{(k)}(x_{\nu})$ provides a representation of the probability density in the space of neutrino fluxes, from which one can evaluate variances, correlations, and higher moments. 
The fits to the FASER 2024 measurements of~\cite{FASER:2024ref} are carried out using $N_{\rm rep}=100$ Monte Carlo replicas.

As customary in the NNPDF approach, fitting is performed by minimizing a quadratic loss function
\begin{equation}
\label{eq:chi-squared}
 E^{(k)} = \frac{1}{n_{\mathrm{dat}}} \sum _{i,j = 1} ^{n_{\mathrm{dat}}} \left( D_i^{(k)} - T_i^{(k)}\right) 
 \lp \mathrm{cov}^{-1}\rp_{ij}  \left( D^{(k)}_j - T_j^{(k)}\right) \, ,
\end{equation}
where the $k$-th theory prediction replica for the $i$-data point is given by
\be
T_i^{(k)} =  \sum_{\alpha=1}^{n_{x}}
 f_{\nu}^{(k)}(x_{\nu,\alpha}){\rm FK}_{\alpha,i} \,, 
\ee
in terms if the $k$-th neural network replica of the neutrino PDF convoluted with the corresponding FK-table.
In Eq.~(\ref{eq:chi-squared}) we use the same covariance matrix as in the Monte Carlo data replica generation of Eq.~(\ref{covmat}).
The overall fit quality is evaluated as
\begin{equation}
\label{eq:chi-squared-v1}
\chi^2 = \frac{1}{n_{\mathrm{dat}}} \sum _{i,j = 1} ^{n_{\mathrm{dat}}} \left( D_i - \la T_i\ra_{\rm rep}\right) 
 \lp \mathrm{cov}^{-1}\rp_{ij}  \left( D_j - \la T_j\ra_{\rm rep} \right) \, ,
\end{equation}
in terms of the average of the theory predictions over the $N_{\rm rep}$ replicas that define the fit ensemble.

In order to fit Eq.~(\ref{eq:NN_parametrisation}) from the binned neutrino event rate (synthetic) data, we use  {\sc\small PyTorch}~\cite{Paszke:2019xhz}.
The neural network architecture in Eq.~(\ref{eq:NN_parametrisation}) is a multi-layer feed-forward perceptron with four hidden layers, each using a softplus activation function to simultaneously enforce a positive output and to add a non-linear component in between layers, such that we use the full complexity of our model.
We adopt the {\sc\small Adam} optimizer with an initial learning rate  \texttt{lr} of 0.01. 
Other training hyperparameters are manually optimized per detector geometry, as the fluxes and the number of data bins differ substantially per setup.\footnote{In future work, an automated hyperparameter optimisation algorithm~\cite{Cruz-Martinez:2024wiu} for the neural net training may be adopted.} 
To avoid overfitting, specially problematic for sparse datasets, it is necessary to introduce a regulation procedure, and here we use  L2 regularisation to constrains model complexity by discouraging large weights by means of a penalty to the loss function.
After training has been successfully completed, the neural network output for the fitted neutrino fluxes is stored in a dense grid in $x_\nu$ compliant with the {\sc\small LHAPDF} format~\cite{Buckley:2014ana}.

\subsection{Closure test validation}
\label{sec:CT}

The NN$\nu$flux fitting methodology is validated by means of closure tests based on synthetic data generated from a known ground truth.
We show that closure tests are successful irrespective of the kinematical variable used as input and of
the detector geometry or integrated luminosity assumed.
We have verified that similar agreement is obtained upon other variation of the fit settings, such as the choice of event generator used for the neutrino flux calculation.
For the Level-2 (L2) closure tests, we generate $N_{\rm L1}=10$ Level-1 (L1) fluctuations around the underlying Level-0 (L0) ground truth for the neutrino event yields and from each of them we generate $N_{\rm rep}=100$ L2 replicas by means of Eq.~(\ref{covmat}), and fit an independent neutrino PDF to each of them.
Subsequently,  the central value and uncertainties of the L2 NN$\nu$flux predictions are computed from the averages and variances, respectively, evaluated over the total $N_{\rm L1}\times N_{\rm rep}=10^3$ fitted replicas.

\begin{figure}[t]
        \centering
\includegraphics[width=0.49\textwidth]{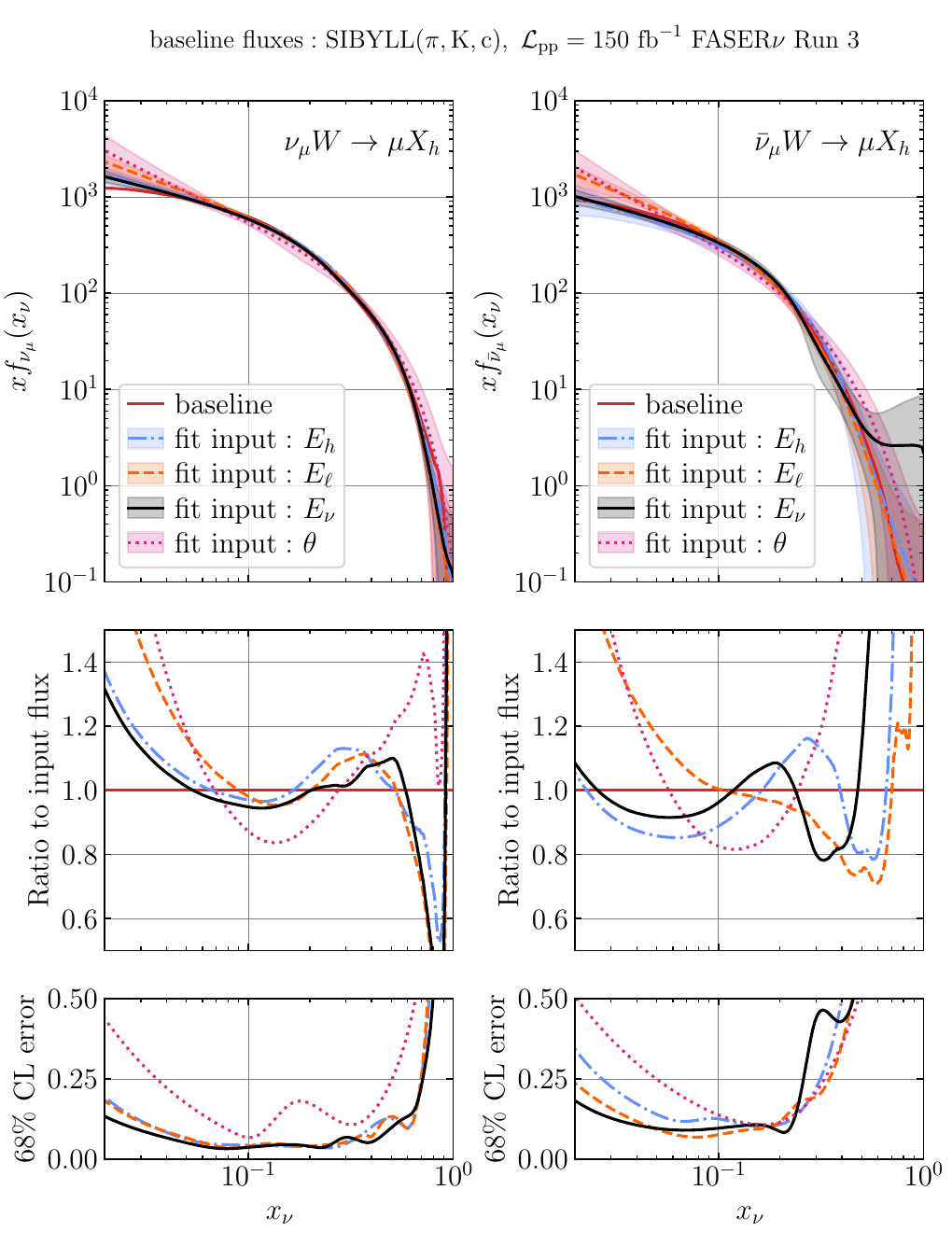}
\includegraphics[width=0.49\textwidth]{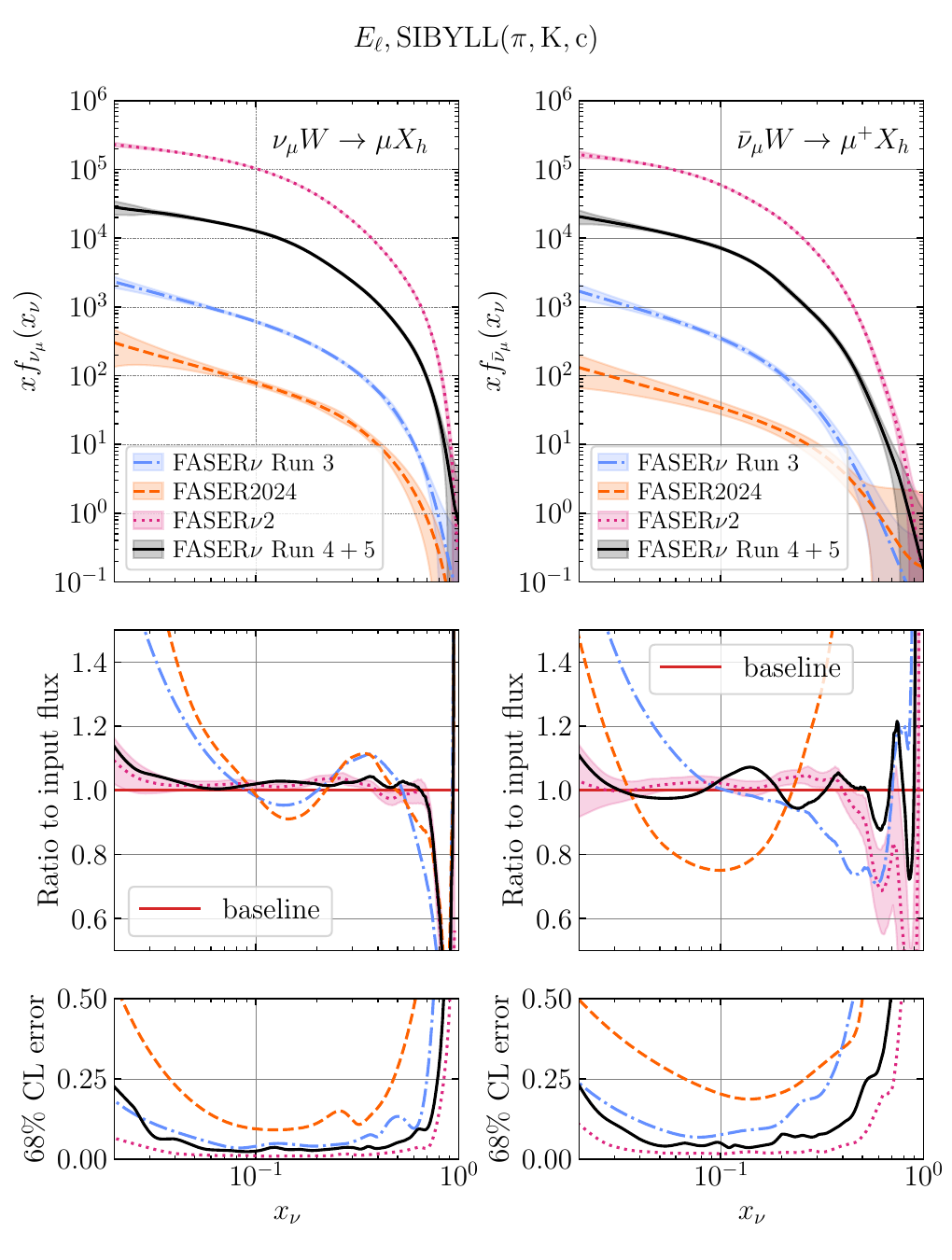}
    \caption{Left: results of NN$\nu$flux L2 closure tests on synthetic data on muon (anti)neutrino scattering generated for four different inputs ($E_h$, $E_\ell$, $E_\nu$, and $\theta_\ell$) for FASER$\nu$ with $\mathcal{L}_{\rm pp}=150$ fb$^{-1}$ using {\sc\small SYBILL}.
    The bands correspond to the fit 68\% CL uncertainties. 
    The middle panel displays the ratio of the NN$\nu$flux fits to the underlying flux, and the bottom one the relative uncertainties.
    Right: same for synthetic data generated for the four detector geometries and luminosities described in Sect.~\ref{eq:overview_fluxes}, for event yields based on the {\sc\small SYBILL} generator binned in  $E_\ell$. 
\label{fig:ct_dependence_on_inputs}
    }
\end{figure}

\paragraph{Stability upon kinematic inputs.}
First we demonstrate that the NN$\nu$flux  methodology closes irrespective of the kinematic input used in the fit. 
The left panels of Fig.~\ref{fig:ct_dependence_on_inputs} displays the results of a L2 closure test on synthetic data generated for FASER$\nu$ with $\mathcal{L}_{\rm pp}=150$ fb$^{-1}$ for neutrino and antineutrino scattering using the {\sc\small SYBILL} event generator both for light and heavy hadron components. 
We compare the ground truth with results of fits based on neutrino event yield data binned in either $E_h$, $E_\ell$, $E_\nu$ or $\theta_\ell$.
The bands correspond to the 68\% CL uncertainties. 
The middle panel displays the NN$\nu$flux results as a ratio to the underlying truth (the {\sc\small SYBILL} muon neutrino flux), and the bottom panel the relative 68\% CL uncertainties in each of the four fits. 

%%%%%%%%%%%%%%%%%%%%%%%%%%%%%%%%
%%%%%%%%%%%%%%%%%%%%%%%%%%%%%%%%

The results of Fig.~\ref{fig:ct_dependence_on_inputs} show that, depite using four different kinematic inputs, the closure test results reproduce in all cases the ground truth at the $1\sigma$ (or at most $2\sigma$) level in the $x_\nu$ region where the bulk of the experimental constraints are located. 
The closure tests are hence successful, in that the true underlying flux is reproduced within uncertainties for most of the $x_\nu$ region and for all four kinematic inputs considered. 
The NN$\nu$fit uncertainties grow rapidly in the small- and large- $x_{\nu}$ extrapolation regions, reflecting the lack of experimental constraints there.
In some cases, specially for muon neutrino scattering, one observes however that the fit results exhibit a different slope than the underlying truth for $x\lsim 0.03$, corresponding to $E_\nu\lsim 100$ GeV.
This is a region where the neutrino event yields are very small and hence results may be affected by the choice of small-$n_\nu$ extrapolation model in Eq.~(\ref{eq:NN_parametrisation}).
This discrepancy is nevertheless not observed  at the level of the neutrino event yields.

The bottom panels of Fig.~\ref{fig:ct_dependence_on_inputs} indicate that, while all four rather kinematic variables satisfy the closure tests, some of them lead to a more precise determination of the neutrino fluxes than others. 
As expected, due to its direct relationship to the input neutrino flux, the $E_\nu$ and $E_\ell$ distributions lead in general to the smallest uncertainties, though fitting $E_h$ enables comparable precision. 
On the other hand, the distribution in the lepton scattering angle $\theta_\ell$ typically displays the largest uncertainties, indicating the looser correlation between this final-state variable and the $E_\nu$ dependence on the initial flux.  

\paragraph{Dependence on detector geometry and luminosity.}
The right panels of Fig.~\ref{fig:ct_dependence_on_inputs} present a similar analysis now for the outputs of NN$\nu$flux L2 closure tests based on synthetic data generated for the four detector geometries and luminosities described in Sect.~\ref{eq:overview_fluxes}:
FASER 2024, FASER$\nu$ at Run 3, FASER$\nu$ at the HL-LHC, and FASER$\nu$2 at the FPF.
In all cases we use event yields binned in $E_\ell$, only statistical errors are considered, and the muon neutrino fluxes are generated with SYBILL both for the light and heavy hadron components. 
These results highlight the increase in neutrino fluxes when either the luminosity or the detector active volume are increased.
The middle panel confirms that in all cases the closure test is satisfied in the kinematic region in $x_\nu$ where there are sufficient experimental constraints, with the same caveat concerning the different small-$x_\nu$ slope as in the comparison of kinematic inputs.  

The bottom panel shows how the uncertainties in the NN$\nu$flux determination decrease as the neutrino event yields of the synthetic data increase: the largest uncertainties are obtained for the FASER 2024 data, while the most precise determination would be obtained with the FASER$\nu$2 detector.
These results indicate that our method appropriately propagates the uncertainties from the input synthetic data to the fitted neutrino fluxes.
As discussed in Sect.~\ref{eq:overview_fluxes}, no attempt is done to estimate the systematic errors and their correlation model, which are expected eventually become the limiting factor for the precision of the analysis~\cite{Cruz-Martinez:2023sdv}.
Accounting for correlated systematic uncertainties in our framework is straightforward following the covariance matrix procedure of Sect.~\ref{sec:nu_flux_fitting}. 

\paragraph{Comparison with synthetic data.}
The analyses of Fig.~\ref{fig:ct_dependence_on_inputs} validate the outcome of closure tests at the level of the fitted neutrino fluxes.
As in the PDF case, one should also validate them at the level of the experimental observables used in the fit, in this case the neutrino scattering binned event rates. 

\begin{figure}[t]
        \centering
\includegraphics[width=0.49\textwidth]{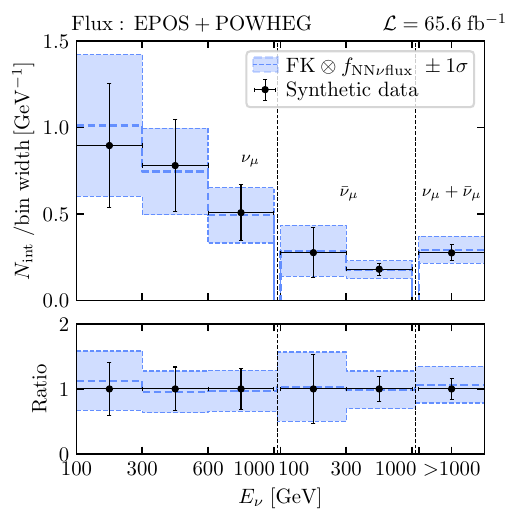}
\includegraphics[width=0.49\textwidth]{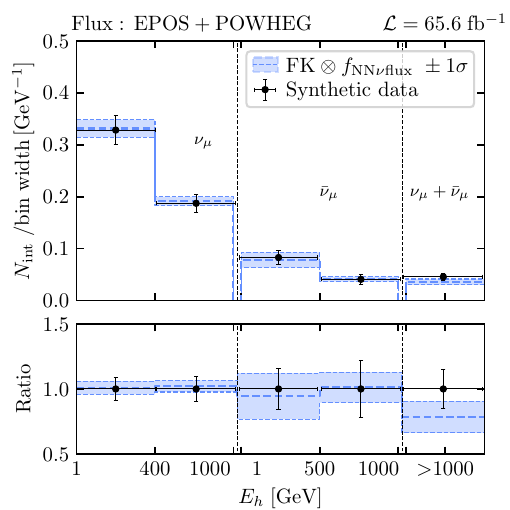}
    \caption{Left: comparison between the results of a NN$\nu$flux L2 closure test fit
    for FASER 2024 with $\mathcal{L}_{\rm pp}=65.6$ fb$^{-1}$ and the corresponding L0 synthetic data binned in $E_\nu$~\cite{FASER:2024ref}.
    The uncertainty band in the NN$\nu$flux predictions corresponds to the 68\% CL uncertainties,
    while the error bars on the L0 synthetic data account both for the statistical and systematic uncertainties as provided in~\cite{FASER:2024ref}. 
    The bottom panel displays the ratio of the fit results to the L0 synthetic data. 
    Right: same for synthetic data binned in $E_h$.
    Note that since the $E_h$ distribution is not measured in~\cite{FASER:2024ref}, here we choose one of the possible binnings and consider only statistical uncertainties. 
    }
\label{fig:closure-test-data-vs-theory-FASERnu150}
\end{figure}

The left panel of
Fig.~\ref{fig:closure-test-data-vs-theory-FASERnu150} displays the results of a NN$\nu$flux L2 closure test carried out for the FASER 2024 (synthetic) data with $\mathcal{L}_{\rm pp}=65.6$ fb$^{-1}$ binned in $E_\nu$, for the same choice of binning as in~\cite{FASER:2024ref}.
The uncertainty band in the NN$\nu$flux predictions corresponds to 68\% CL interval evaluated from the $N_{\rm L1}\times N_{\rm rep}=10^3$ fitted neutrino PDF replicas, while the error bars on the L0 synthetic data are the sum in quadrature of the statistical and systematic uncertainties as estimated in~\cite{FASER:2024ref}.
The bottom panel displays the ratio of the fit results to the central value of the L0 observables.
From Fig.~\ref{fig:closure-test-data-vs-theory-FASERnu150} we observe agreement between the NN$\nu$flux closure test results and the input L0 synthetic data for both central values of uncertainties.
As expected due to the broad $E_\nu$ bins, the NN$\nu$flux fit uncertainties are similar to those of the synthetic data. 

%%%%%%%%%%%%%%%%%%%%%%%%%%%%%%%%
%%%%%%%%%%%%%%%%%%%%%%%%%%%%%%%%%%%%

The right panel of
Fig.~\ref{fig:closure-test-data-vs-theory-FASERnu150} presents a similar comparison now for the FASER 2024 synthetic data binned in $E_h$.
Since this $E_h$ distribution was not measured in~\cite{FASER:2024ref}, here we choose one among the possible binnings and consider only statistical uncertainties from the expected event yields in each bin.
Therefore, differences in the L0 synthetic data uncertainties between the $E_\nu$ and $E_h$ closure tests stem both from the lack of systematic errors in the latter as well as from the different binning (the total integrated event yields are the same in the two distributions).
Also for this distribution the closure test is successful, and the reduced uncertainties in the NN$\nu$fit  propagate those of the input synthetic data.

\begin{figure}[t]
        \centering
\includegraphics[width=0.65\textwidth]
{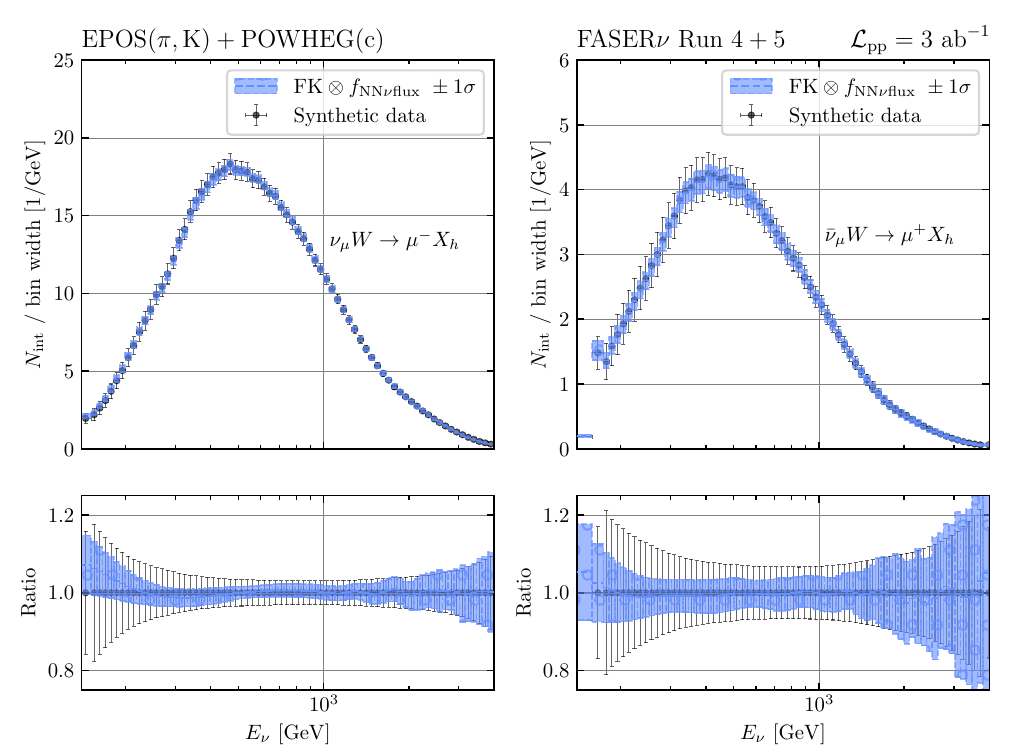}
\includegraphics[width=0.65\textwidth]
{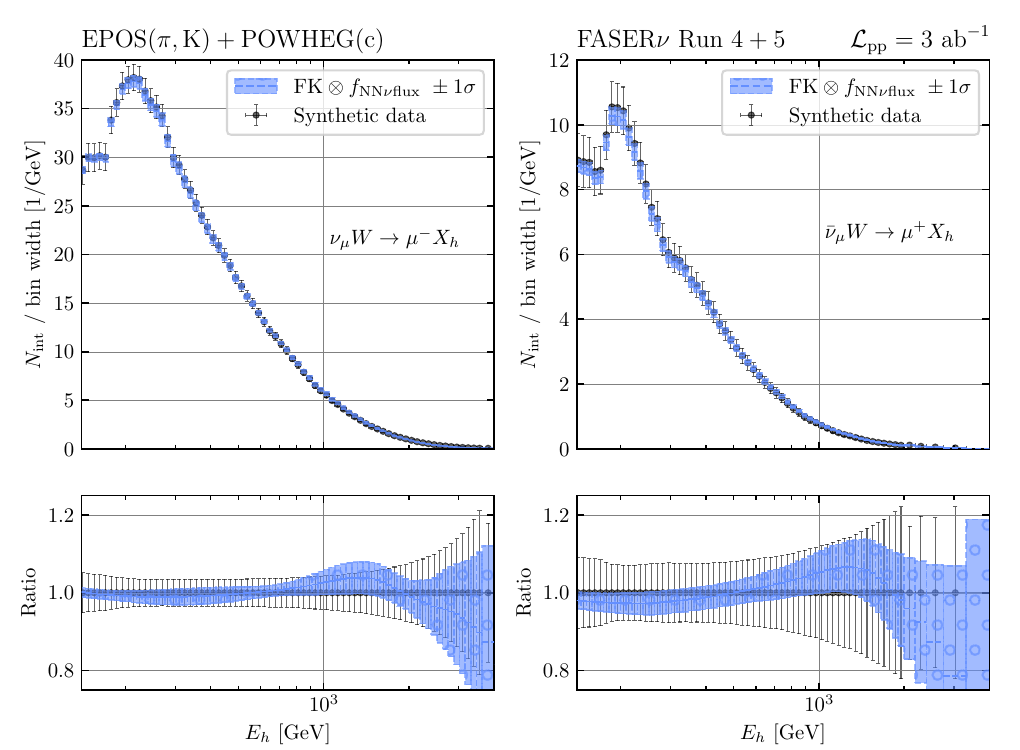}
    \caption{Same as Fig.~\ref{fig:closure-test-data-vs-theory-FASERnu150} for muon neutrino and antineutrino scattering  at FASER$\nu$ at the HL-LHC for synthetic data binned in $E_\nu$ (top) and $E_h$ (bottom panels).  
    }
\label{fig:closure-test-data-vs-theory-FASER2HLLHC}
\end{figure}

In the same manner as in Fig.~\ref{fig:ct_dependence_on_inputs}, we 
also verify that  closure tests are fulfilled for larger detectors and for higher integrated luminosities.
To illustrate this, Fig.~\ref{fig:closure-test-data-vs-theory-FASER2HLLHC} presents L2 closure tests results carried out with the same settings as in  Fig.~\ref{fig:closure-test-data-vs-theory-FASERnu150} now for FASER$\nu$ at the HL-LHC with $\mathcal{L}_{\rm pp}=3$ ab$^{-1}$.
We present results both  for the $E_\nu$ and $E_h$ distributions, separating neutrino and antineutrino scattering event yields, and only statistical uncertainties are considered in the synthetic L0 data.
Good agreement with the input  data is obtained in all cases for the complete kinematic range of $E_\nu$ and $E_h$ considered.
At odds with the FASER 2024 closure tests of Fig.~\ref{fig:closure-test-data-vs-theory-FASERnu150},  the NN$\nu$flux prediction is now more precise that the synthetic data in the bulk of the distribution, likely due to the correlated information provided by neighbouring bins.
In the large-$x_\nu$ extrapolation regions instead, where event yields are suppressed, the NN$\nu$flux uncertainties match those of the L0 synthetic data. 

%%%%%%%%%%%%%%%%%%%%%%%%%%%%%%%%
%%%%%%%%%%%%%%%%%%%%%%%%%%%%%%%%%%%%

\paragraph{Additional checks.}
Following the NNPDF closure test framework, we have verified that the NN$\nu$flux uncertainty estimated in L2 closure tests is independent of the specific L1 instance in the data region, while larger variations are found in the extrapolation regions.
We have also verified that the distribution of loss functions $\chi^2$, Eq.~(\ref{eq:chi-squared-v1}), over a large enough sample of L2 closure tests, each one based on a different synthetic L1 instance, follow the expected Gaussian distribution.

\section{The muon neutrino flux from FASER 2024 data}
\label{sec:results_faser}

Now we apply the NN$\nu$flux methodology presented in Sect.~\ref{sec:formalism} to carry out a first determination of the LHC muon neutrino fluxes from the FASER 2024 measurement of~\cite{FASER:2024ref}.
In this analysis, the covariance matrix used in the Monte Carlo replica generation Eq.~(\ref{covmat}) and in the fit loss function Eq.~(\ref{eq:chi-squared}) accounts both for the statistical and for the experimental systematic uncertainties from~\cite{FASER:2024ref}.

\begin{figure}[t]
        \centering
\includegraphics[width=0.48\textwidth]{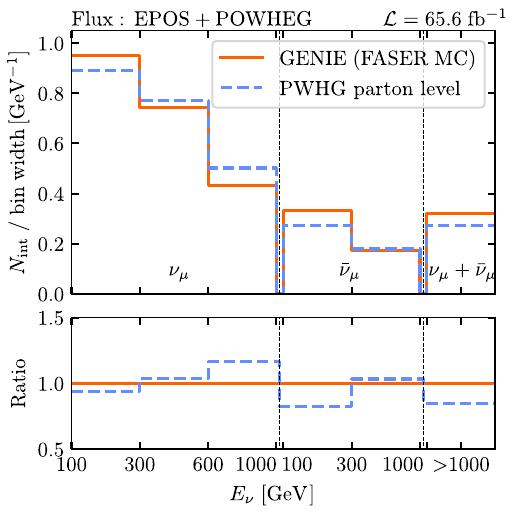}
\includegraphics[width=0.48\textwidth]{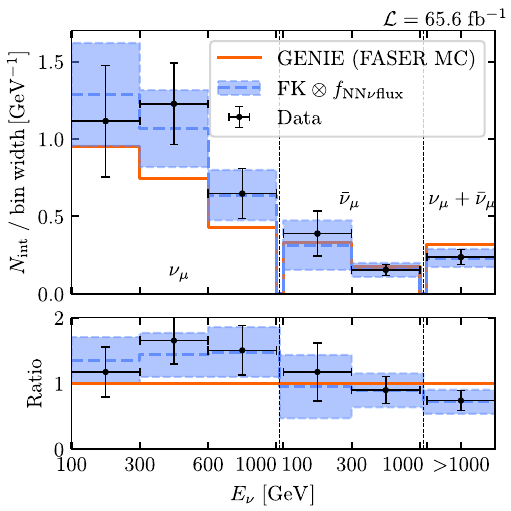}
\caption{Left: the predicted number of muon neutrino interactions (normalised by bin width) at FASER for $\mathcal{L}_{\rm pp}=65.6$ fb$^{-1}$ and obtained with {\sc\small POWHEG~DIS} event generator compared to its counterpart used in the FASER publication~\cite{FASER:2024ref} and based on the {\sc\small GENIE} generator.
The same neutrino fluxes, {\sc\small EPOS+POWHEG}, are used in both calculations.
The bottom panel displays the ratio between the two  predictions.
Right: comparison between the NN$\nu$flux fit results for the muon neutrino flux and the FASER 2024 data.
We also display the {\sc\small GENIE} predictions based on the {\sc\small EPOS+POWHEG} muon neutrino flux.
The bottom panel displays the ratio to the central value of the data. 
The uncertainties in the fit predictions correspond to the 68\% CL intervals.
}
\label{fig:MC_unfolded_comparison}
\label{fig:FASERv_fit_comparison_data}
\end{figure}

First we assess how the {\sc\small POWHEG+Pythia8} NLO+LL pipeline used in this work compares with the  {\sc\small GENIE}-based predictions used in the FASER analysis. 
The left panel of Fig.~\ref{fig:MC_unfolded_comparison} displays the predicted number of muon neutrino interactions, normalised by bin width, as a function of $E_\nu$ at FASER for $\mathcal{L}_{\rm pp}=65.6$ fb$^{-1}$ and obtained with the {\sc\small POWHEG~DIS}  framework, compared with the {\sc\small GENIE} predictions taken from~\cite{FASER:2024ref}.
The same input muon neutrino flux is used in the two calculations. 
From Fig.~\ref{fig:MC_unfolded_comparison} one observes that differences between the {\sc\small POWHEG~DIS} and {\sc\small GENIE} simulations range between a few percent and at most $\sim 20\%$, consistently with the findings  of~\cite{vanBeekveld:2024ziz,Candido:2023utz}.
We have also verified that integration uncertainties are negligible (sub-permille) in the {\sc\small POWHEG~DIS} predictions.

One major difference between the {\sc\small POWHEG~DIS} and {\sc\small GENIE} calculations is the neutrino DIS structure function model, with the latter using Bodek-Yang~\cite{Bodek:2002vp}  which neglects (N)NLO QCD corrections and is based on the GRV98 LO PDFs as opposed to PDF4LHC21 NNLO in {\sc\small POWHEG~DIS}.
Furthermore, {\sc\small GENIE} includes contributions due to non-DIS processes, such as shallow inelastic scattering and resonance production, which can account for about $5\%$ to $10\%$ of the cross section for neutrino energies with $E_\nu \sim 100$~GeV.
Finally, we note that the {\sc\small GENIE} and {\sc\small POWHEG~DIS} simulations of event yields may be affected by other differences in addition to the neutrino cross-section model, which can only be resolved by means of a dedicated benchmark. 

%%%%%%%%%%%%%%%%%%%%%%%%%%%%%%%%
%%%%%%%%%%%%%%%%%%%%%%%%%%%%%%%%%%%%%%

\begin{figure}[h]
        \centering
\includegraphics[width=0.80\textwidth]{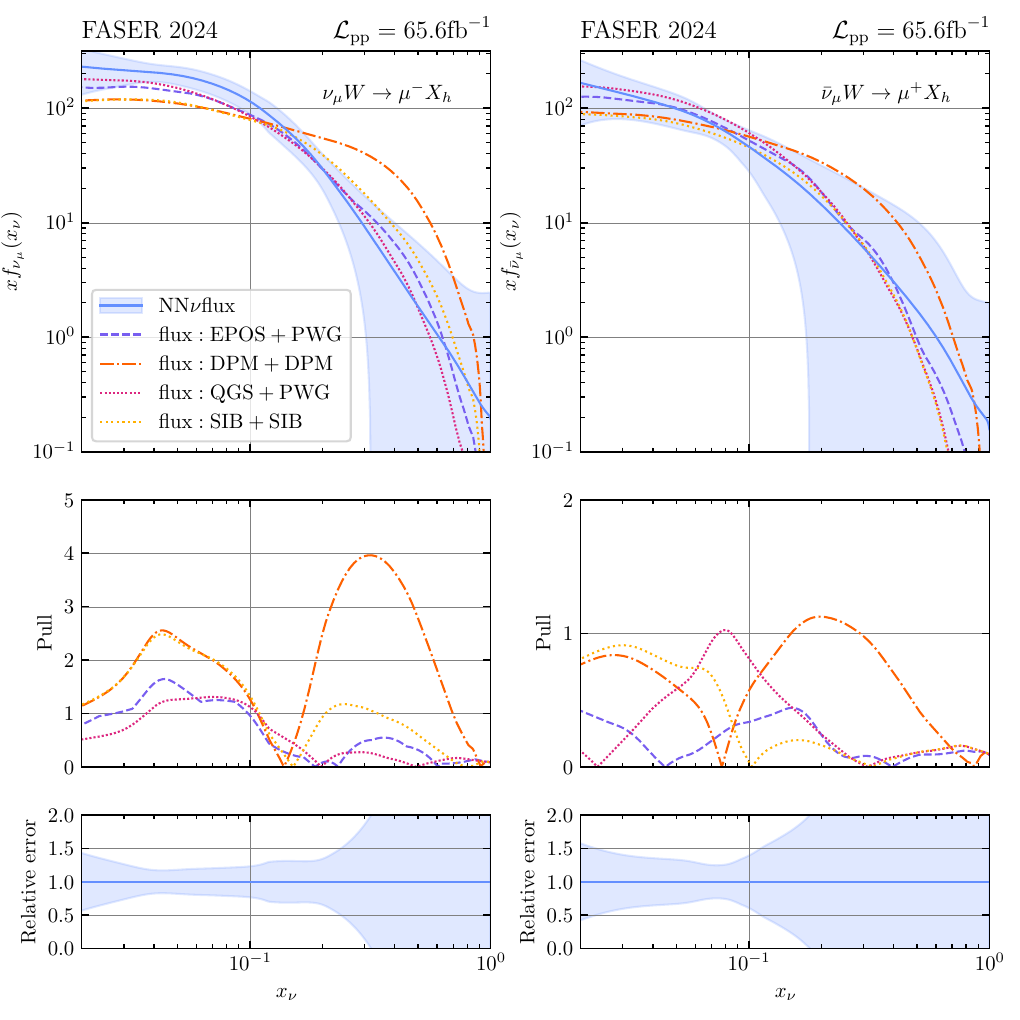}
\caption{The NN$\nu$flux determination of the muon neutrino and antineutrino fluxes from the FASER 2024 measurements of the event yields binned in $E_\nu$.
The bands in the fit results indicate the 68\% CL uncertainties.  
The NN$\nu$flux determination  is compared with the predictions based on four different neutrino fluxes: {\sc\small EPOS+POWHEG}, {\sc\small DPMJET+DPMJET}, {\sc\small QGSJET+POWHEG}, and {\sc\small SIBYLL+SIBYLL}, where the first (second) generator simulates light (heavy) hadron production.
The middle panel indicates the pull between the flux predictions and the NN$\nu$flux results, Eq.~(\ref{eq:pull_FASERdata}), while the bottom panel display the relative uncertainties of the NN$\nu$flux determination.
See the right panel Fig.~\ref{fig:FASERv_fit_comparison_data} for the corresponding comparison with the FASER 2024 data.
\label{fig:FASER_sigma_deviation_savgol}
    }
\end{figure}

Moving to the  fit results, Fig.~\ref{fig:FASER_sigma_deviation_savgol} displays the results of the NN$\nu$flux determination of the muon neutrino and antineutrino PDFs from the FASER 2024 data binned in $E_\nu$, where the band indicates the 68\% CL uncertainties from the fit. 
The NN$\nu$flux results are compared to the predictions based on the four different fluxes as discussed in Sect.~\ref{eq:overview_fluxes}: {\sc\small EPOS+POWHEG}, {\sc\small DPMJET+DPMJET}, {\sc\small QGSJET+POWHEG}, and {\sc\small SIBYLL+SIBYLL}, where the first (second) generator simulates light (heavy) hadron production. 
The middle panel indicates the pull between each of the flux predictions and the NN$\nu$flux results, defined as
\be
\label{eq:pull_FASERdata}
P_{\rm gen}(x_\nu) = \frac{\Big|\la f^{(\rm {\rm nn}\nu{\rm fit})}_{\nu_\mu}(x_\nu)\ra- f^{(\rm gen)}_{\nu_\mu}(x_\nu)\Big|}{\sigma^{{\rm nn}\nu{\rm fit}}_{\nu_\mu} (x_\nu)
} \, ,
\ee
where $\langle f^{(\rm {\rm nn}\nu{\rm fit})}_{\nu_i}\rangle$ and $\sigma^{{\rm nn}\nu{\rm fit}}_{\nu_\mu}$ indicate the central value and uncertainties of the fit results and $f^{(\rm gen)}_{\nu_\mu}$ corresponds to the prediction from the four generators considered.

%%%%%%%%%%%%%%%%%%%%%%%%%%%%%%%%
%%%%%%%%%%%%%%%%%%%%%%%%%%%%%%%%

From  Fig.~\ref{fig:FASER_sigma_deviation_savgol} one finds that the uncertainties in the NN$\nu$flux determination are around 30\% in the data region, consistent with the experimental precision of the inputs, and then grow large in the high-$x_\nu$ region due to the lack of constraints. 
From the pulls between the NN$\nu$flux fit and the predictions for the various event generators, the FASER data exhibits clear discrimination power. 
Specifically, the DPMJET predictions are strongly disfavoured, at the $4\sigma$ level, in the region $x_\nu \gsim 0.2$ (corresponding to $E_\nu \gsim 1$ TeV), where much higher rates of neutrinos and antineutrinos are predicted than those observed by FASER.
The SYBILL prediction is also disfavoured, overshooting the NN$\nu$flux fit in the multi-TeV region and undershooting it for smaller $x_\nu$ values, as is also the case with DPMJET. 
The flux predictions based on either EPOS or QGSJET for light hadron production and on POWHEG for charm production agree better with the NN$\nu$flux fit, with pulls below $2\sigma$ level in the full energy range. 
We study in Sect.~\ref{subsec:generators} how this discrimination power between predictions from event generators for forward hadron production at FASER is further enhanced for electron neutrinos and for higher-statistics data-taking periods. 

The right panel of Fig.~\ref{fig:FASERv_fit_comparison_data}  presents the corresponding comparison between the NN$\nu$flux results for the muon neutrino event yields binned in $E_\nu$ and the FASER 2024 data~\cite{FASER:2024ref}, where the error bars are the sum in quadrature of the statistical and systematic uncertainties.
For reference, we also display the same {\sc\small GENIE} predictions from~\cite{FASER:2024ref} as in the left panel, based on {\sc\small EPOS-LHC} for light and {\sc\small POWHEG} for charm hadron production.  
Results are presented in 6 bins of $E_\nu$ and muon charge $q$, and the  fit uncertainties correspond to 68\% CL intervals.
The bottom panel displays the ratio to the central value of the data. 
This comparison highlights the improved agreement of NN$\nu$flux with the data as compared to the muon neutrino flux ({\sc\small EPOS+POWHEG}) adopted in~\cite{FASER:2024ref}.
Specifically, the fitted NN$\nu$flux muon neutrino flux is larger than the {\sc\small EPOS+POWHEG} predictions by between 20\% and 40\%, although both fluxes agree within the fit uncertainties at the $1\sigma$ level. 
The NN$\nu$flux predictions exhibit similar precision to that of the FASER data, consistent with the fact that the broad bins in $E_\nu$ constrain independent $x_\nu$ regions of the fluxes.

Figs.~\ref{fig:FASERv_fit_comparison_data} and~\ref{fig:FASER_sigma_deviation_savgol} contain the main results of this work: a first extraction of the LHC forward neutrino fluxes from the data taken at the LHC far-forward neutrino detectors.
In addition to providing a proof-of-concept of our methodology, these results demonstrate that FASER already has competitive sensitivity for QCD studies such as discriminating between event generators of forward hadron production.
In Sect.~\ref{sec:interpretation} we extend these studies to projections for future data-taking periods and to other representative applications. 

\section{Applications to LHC forward particle production}
\label{sec:interpretation}

Here we present three representative applications of the NN$\nu$flux methodology for the determination of the LHC forward neutrino fluxes.
First, following up on the initial discussion of Sect.~\ref{sec:results_faser}, we assess how electron and muon neutrino flux extractions can validate or disfavour Monte Carlo event generators of forward hadron production.
Second, we quantify to which extent fits of the electron and muon neutrino fluxes can constrain the charm quark content of the proton and isolate a possible intrinsic component.
Third, we test beyond the SM (BSM) scenarios displaying an increased branching ratio of neutral hadrons into neutrinos and show that FASER can provide stringent constraints which significantly improve the current experimental bounds.

\subsection{Benchmarking forward event generators}
\label{subsec:generators}

%%%%%%%%%%%%%%%%
\begin{figure}[t]
\centering
\includegraphics[width=0.43\textwidth]{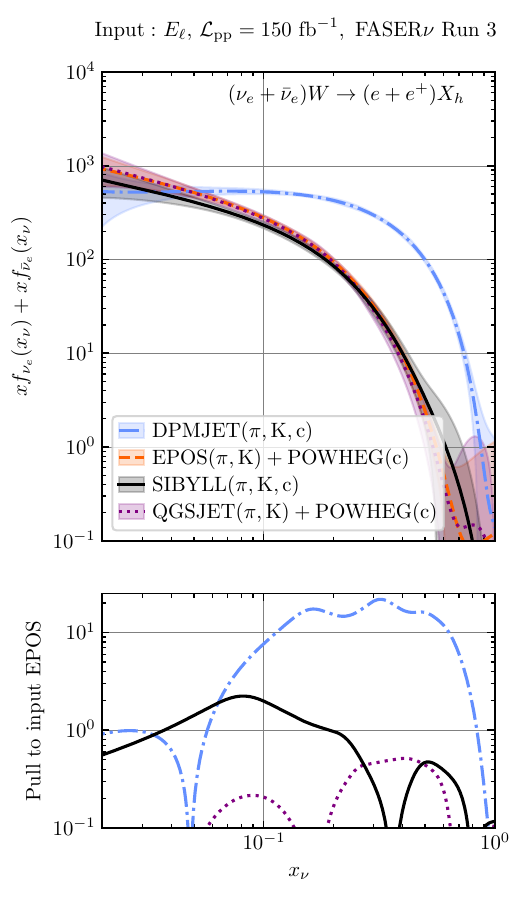}
\includegraphics[width=0.43\textwidth]{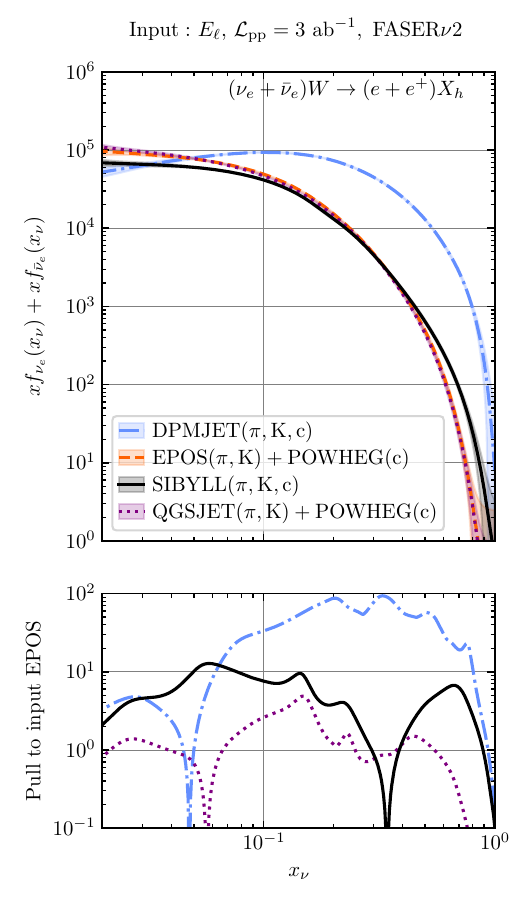}
    \caption{NN$\nu$flux fits to synthetic data for electron neutrino and antineutrino scattering binned in $E_\ell$ and generated using four different electron neutrino flux predictions: {\sc\small DPMJET}, {\sc\small EPOS+POWHEG}, {\sc\small SIBYLL}, and {\sc\small QGSJET+POWHEG}.
    We present results both for
    FASER$\nu$ with $\mathcal{L}_{\rm pp}=150$ fb$^{-1}$ (left) and FASER$\nu$2 with $\mathcal{L}_{\rm pp}=3$ ab$^{-1}$ (right panel).
    The bottom panels display the pull Eq.~(\ref{eq:fluxes_comparison_generators}) with respect to the baseline fit based on the {\sc\small EPOS+POWHEG} event yields.
    Note the logarithmic scale in the $y$ axes of the pull panels.
    }
\label{fig:comparison_generators_electron}
\end{figure}
%%%%%%%%%%%%%%%%%%

In Fig.~\ref{fig:FASER_sigma_deviation_savgol} we compared the NN$\nu$flux muon neutrino fit to the FASER 2024 data to the predictions based on different event generators for forward hadron production and observed a marked discrimination power.
We now extend that analysis to electron neutrino event yields and other data-taking periods and detectors.
Fig.~\ref{fig:comparison_generators_electron} displays the results of the NN$\nu$flux fits to synthetic data on electron neutrino event yields, binned  in $E_\ell$ and generated using four different electron neutrino flux predictions: {\sc\small DPMJET}, {\sc\small EPOS+POWHEG}, {\sc\small SIBYLL}, and {\sc\small QGSJET+POWHEG}.
The bottom panels display the pull with respect to the baseline fit:
\be
\label{eq:fluxes_comparison_generators}
P_{\rm gen}(x_\nu) = \frac{\Big|\la f^{(\rm gen)}_{\nu_i}(x_\nu)\ra-\la f^{(\rm gen0)}_{\nu_i}(x_\nu)\ra\Big|}{\sqrt{
\sigma^{(\rm gen)2}_{\nu_i}+   \sigma^{(\rm gen0)2}_{\nu_i} 
}} \, ,
\ee
where gen0 is the result of the fit based on the EPOS+POWHEG synthetic data and $\sigma^{(\rm gen)}_{\nu_i}$ is the 68\% CL uncertainty of the fit.
Results are presented for the sum over electron neutrino and antineutrino fluxes for FASER$\nu$ at Run 3 and FASER$\nu$2 at the FPF. 

The pull analysis in the bottom panels of Fig.~\ref{fig:comparison_generators_electron} further corroborate the findings of Sect.~\ref{sec:results_faser} concerning the sensitivity of FASER to discriminate between the theoretical predictions based on different event generators. 
For FASER$\nu$ at Run 3, large pulls for {\sc\small DPMJET} are observed ($P_{\rm gen}\gsim 10$), while $P_{\rm gen}\lsim 2$ for the other generators.
Instead, for FASER$\nu$2 at the FPF, the increased data statistics lead to an extreme discrimination power, and we find that predictions from the four event generators considered can be disentangled at the $5\sigma$ level or higher in all cases.
This result highlights the unique FASER sensitivity to constrain models of forward particle production in hadronic collisions, which in turn enables more reliable prediction for high-energy cosmic ray and astroparticle physics experiments. 

\begin{figure}[t]
\centering
\includegraphics[width=0.86\textwidth]{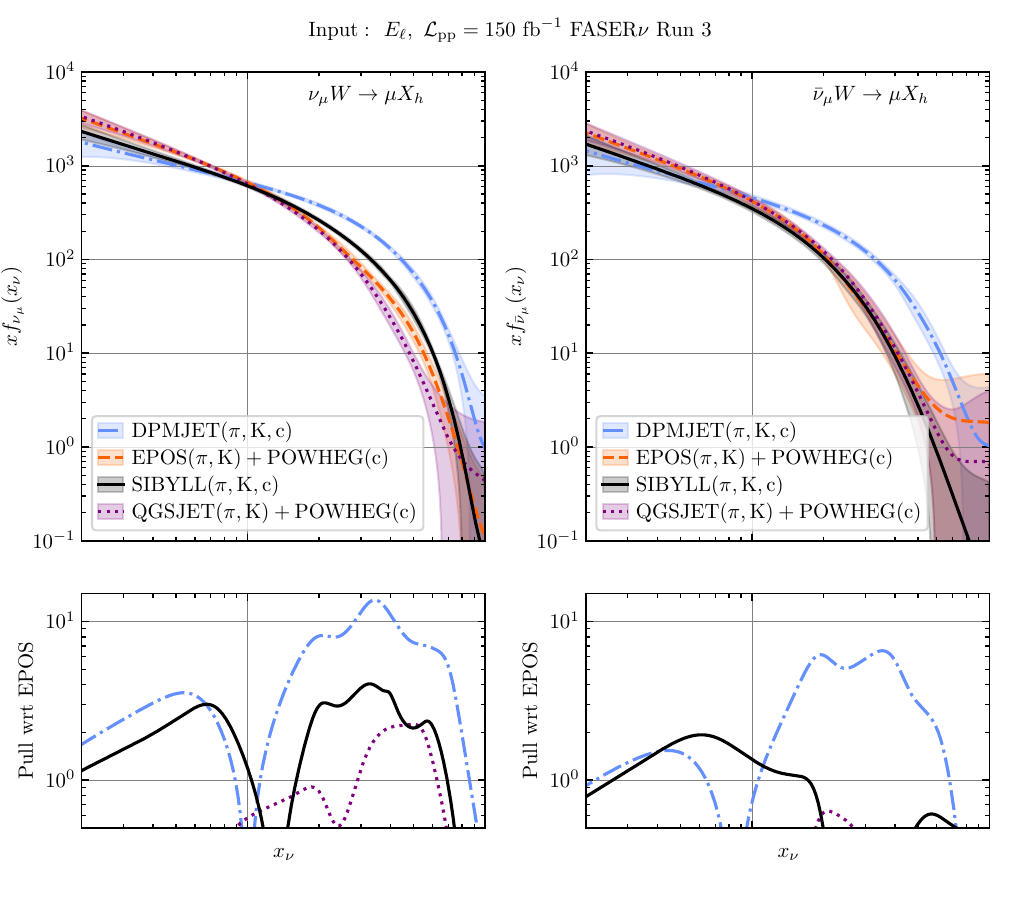}
    \caption{Same as Fig.~\ref{fig:comparison_generators_electron} for synthetic data on muon neutrino (left) and antineutrino (right panel) event yields for FASER$\nu$ at Run 3.
    }
\label{fig:comparison_generators_muon}
\end{figure}

The same comparison as in Fig.~\ref{fig:comparison_generators_electron} is shown in Fig.~\ref{fig:comparison_generators_muon} for muon neutrino event yields at FASER$\nu$ Run 3, where we display separately the NN$\nu$flux fit results for neutrinos and antineutrinos.
A similar picture arises, with differences between the DPMJET and EPOS+POWHEG reaching above the $10\sigma$ level for muon neutrino scattering.
We note that, in general, the pull pattern is similar for muon neutrinos and antineutrinos, and thus the separate measurement of the neutrino and antineutrino event yields provides an internal validation of the analysis interpretation in terms of models for forward hadron production. 

%%%%%%%%%%%%%%%%
%%%%%%%%%%%%%%%%%%

The results from Figs.~\ref{fig:comparison_generators_electron} and~\ref{fig:comparison_generators_muon} demonstrate that neutrino flux predictions based on {\sc\small DPMJET} are very different from those provided by the other generators considered.
As already noted in~\cite{Kling:2021gos}, this is because {\sc\small DPMJET} predicts a significantly increased rate of forward charm production: its resulting neutrino flux from charm decays exceeds that of other predictions by more than a factor 10.
As studied in detail in~\cite{FASER:2024ykc}, the main reason for this feature is the fact that DPMJET neglects charm mass effects when calculating the underlying hard scattering matrix element, which are essential to correctly describe the kinematics of forward $D$-meson production. 
In addition, DPMJET uses the CT14LO PDFs~\cite{Dulat:2015mca}, which may overestimate the charm quark content, as well as a constant $K$-factor factor $\sim$2 to model higher-order QCD corrections that, although
not unreasonably large in this context, significantly enhances the rate.
We note that disagreements with the DPMJET predictions have also been observed for LHCb open charm measurements~\cite{LHCb:2015swx}. 

Figs.~\ref{fig:comparison_generators_electron} and~\ref{fig:comparison_generators_muon} also illustrate the  constraining power of FASER and its upgrades on hadronic interaction models describing forward particle production at high energies, which play a crucial role in many astroparticle physics applications. 
Indeed, hadronic interaction models are used to describe both particle production in extreme astrophysical environments and cosmic ray interactions in Earth’s atmosphere. 
Notably, in the latter case, there is a long-standing discrepancy --- commonly referred to as the muon puzzle --- between the number of muons observed in high-energy cosmic ray air showers and predictions from hadronic models~\cite{PierreAuger:2014ucz, PierreAuger:2016nfk, EAS-MSU:2019kmv, Soldin:2021wyv, PierreAuger:2024neu}.
This mismatch currently hampers efforts to determine the mass composition of cosmic rays and to discriminate between different hypotheses about their origins. 
Extensive studies over the past decade suggest the discrepancy likely stems from a mis-modeling of soft QCD effects in forward particle production at center-of-mass energies above the TeV scale~\cite{Ulrich:2010rg, Albrecht:2021cxw}, such as an underestimation of forward strangeness production~\cite{Allen:2013hfa, Anchordoqui:2016oxy, Anchordoqui:2019laz}. 
As already discussed, LHC neutrino flux measurements can constrain these models and test proposed explanations for the muon puzzle~\cite{Anchordoqui:2022fpn, Sciutto:2023zuz}.

Another area where LHC neutrino flux measurements can benefit astroparticle physics is the prompt atmospheric neutrino flux, which arises from the decay of charmed hadrons produced in cosmic ray collisions with the atmosphere. 
At neutrino energies above a few hundred TeV, this flux constitutes the dominant background in searches for astrophysical neutrinos by telescopes such as IceCube~\cite{Abbasi:2021qfz}. 
However, its magnitude is currently subject to large uncertainties, which limit the extraction of the astrophysical neutrino flux. 
Collider neutrino flux measurements, which indirectly probe forward charm production at the LHC, can help reduce these uncertainties and enhance the sensitivity of neutrino observatories and multi-messenger astronomy~\cite{Bai:2022xad}.
This potential is also illustrated in the next subsection, which studies the potentially large effects of charm-initiated processes in forward $D$-meson production.

\subsection{Intrinsic charm in forward $D$-meson production}
\label{subsec:application_IC}

At hadron colliders such as the LHC, the production of charmed hadrons is dominated by the $g+g\to c+\bar{c}$ channel, with subleading contributions from quark-initiated processes such as $c + g \to c+X$ being often neglected. 
While ignoring this charm-gluon initial-state mechanism  represents an excellent approximation for PDF sets where charm is generated from perturbative radiation off gluons and light quarks, it does not necessarily hold true for PDF sets where charm is fitted~\cite{Guzzi:2022rca,NNPDF:2021njg,Ball:2022qks} and which hence support an intrinsic charm component~\cite{Brodsky:1980pb, Hobbs:2017fom, Hobbs:2013bia,Brodsky:2015fna}.
The contribution from the $gc$ partonic luminosity in PDF sets allowing for intrinsic charm (IC) is specially important in the forward region sensitive to large-$x$ PDFs~\cite{LHCb:2021stx,LHCb:2022cul}.
Indeed, in the forward region, $D$-meson production from gluon-charm scattering becomes sizeable in scenarios with IC and eventually dominates over the gluon-fusion channel~\cite{Maciula:2022lzk}.  
Therefore, as well known, forward neutrino fluxes from $D$-meson decays provide a sensitive probe of the intrinsic charm content of the proton.

%%%%%%%%%%%%%%%%
\begin{figure}[t]
\centering
\includegraphics[width=0.42\textwidth]{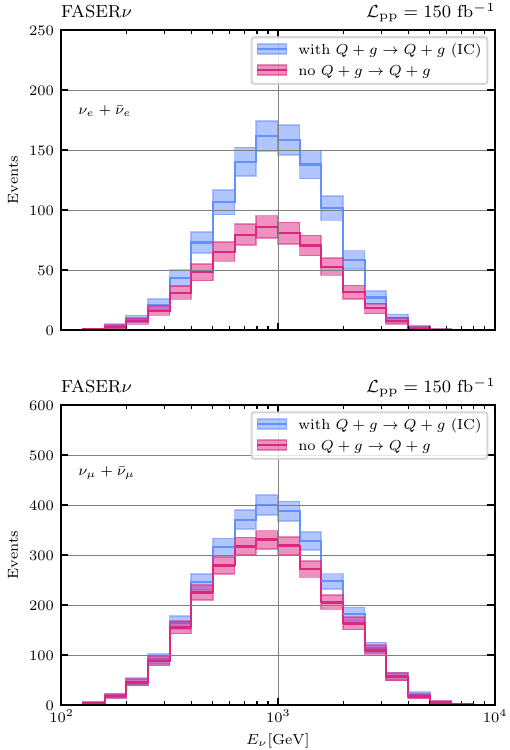}
\includegraphics[width=0.42\textwidth]{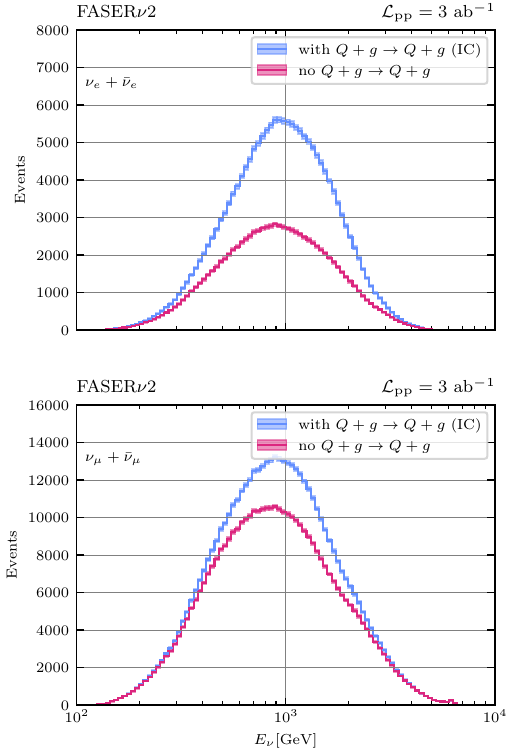}
    \caption{The predicted number of interacting electron (top)  and muon (bottom panels) neutrino+antineutrino  events
    at FASER$\nu$ with $\mathcal{L}_{\rm pp}=150$ fb$^{-1}$ (left) and at  FASER$\nu$2 with $\mathcal{L}_{\rm pp}=3$ ab$^{-1}$ (right panels) as a function of $E_\nu$.
    The baseline flux prediction is the {\sc\small EPOS} (light hadrons) + {\sc\small POWHEG} (heavy hadrons) calculation.
    The  {\sc\small POWHEG} charm production calculation is extended with the contribution from the $g+Q \to Q +X$ production channel from~\cite{Maciula:2022lzk} based on CT14 IC BHPS as input PDF set.
    The bands in each bin correspond to the statistical uncertainties.
    }
\label{fig:IC_events_comp}
\end{figure}
%%%%%%%%%%%%%%%%%%

Extending the NLO {\sc\small POWHEG}~\cite{Alioli:2010xd} calculation of charm production and decay into neutrinos of Ref.~\cite{Buonocore:2023kna} to the intrinsic charm case is complicated by the fact that {\sc\small POWHEG} works in a $n_f=3$ fixed flavour number (FFN) scheme where charm is absent from the initial state.
While calculations of NLO charm production in general-mass schemes~\cite{Cacciari:2012ny,Kniehl:2012ti,Helenius:2018uul} combining the massive ($n_f=3$) and massless ($n_f=4$) calculations are available, they need to be modified to account for possible intrinsic charm contributions, as was done for DIS in~\cite{Ball:2015tna}.
Here we take a simpler approach, and add the calculation of forward neutrino production in the $g+c$ channel using $k_T$ factorisation from Ref.~\cite{Maciula:2022lzk} to the {\sc\small POWHEG} FFN predictions of Ref.~\cite{Buonocore:2023kna}.
This approximation,  justified at the synthetic data level, enables us to obtain a first quantitative estimate of the sensitivity of FASER to intrinsic charm effects in forward $D$-meson production.
When confronting  with experimental data, however, one should instead use a consistent (N)NLO matched calculation accounting for the charm PDF contribution in the initial state of the reaction, and also account for theoretical uncertainties such as missing higher orders and $m_c$ variations. 

Fig.~\ref{fig:IC_events_comp} displays the predicted number of interacting electron and muon  neutrinos and antineutrinos at FASER$\nu$ for $\mathcal{L}_{\rm pp}=150$ fb$^{-1}$  and at FASER$\nu$2 for $\mathcal{L}_{\rm pp}=3$ ab$^{-1}$, respectively, as a function of $E_\nu$.
The baseline flux prediction is based on EPOS for light hadron and on POWHEG for heavy hadron production.
We extend the latter with the contribution from the  $g+Q \to Q +X$ channel from Ref.~\cite{Maciula:2022lzk}, based on the CT14 IC set~\cite{Hou:2017khm} for the BHPS model~\cite{Brodsky:1980pb}, in an scenario where intrinsic charm carries $\sim 1\%$ of the total proton momentum (see also~\cite{Guzzi:2022rca} for more recent CT studies with fitted charm PDFs).
In the case of electron neutrinos, accounting for the $gQ$ partonic luminosity with intrinsic charm in the charm production predictions leads to an increase in the expected event yields of by up to a factor 2 starting from $E_\nu\sim 0.5$ TeV.
For the muon neutrino event yields, the relative effects of charm-initiated contributions are smaller, as expected from the subdominant role played by the contribution from $D$-mesons as compared to light hadrons in muon neutrino production~\cite{Kling:2021gos,FASER:2024ykc}.
Therefore, in the following we focus on the NN$\nu$flux fits based on synthetic data to the electron neutrino event yields to assess whether FASER measurements can fingerprint IC in the proton.

%%%%%%%%%%%%%%%%%%%%%%%%%%
\begin{figure}[t]
        \centering
\includegraphics[width=0.42\textwidth]{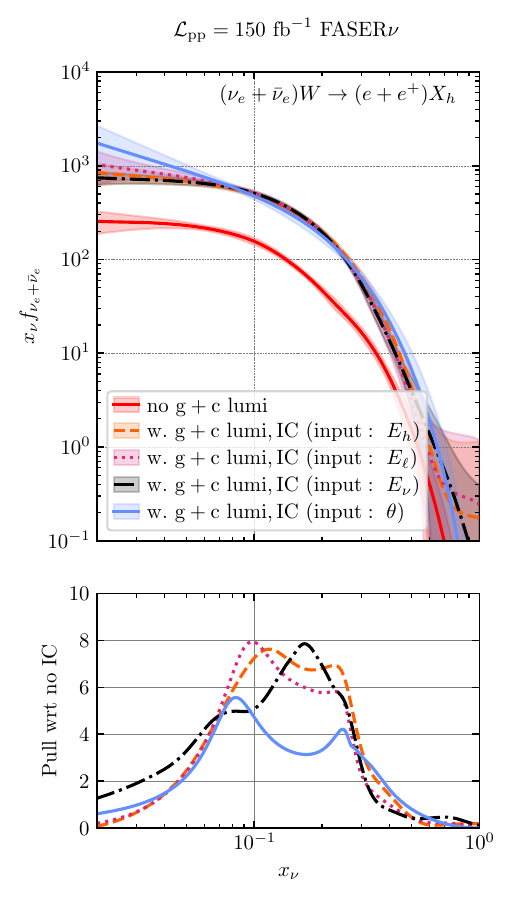}
\includegraphics[width=0.42\textwidth]{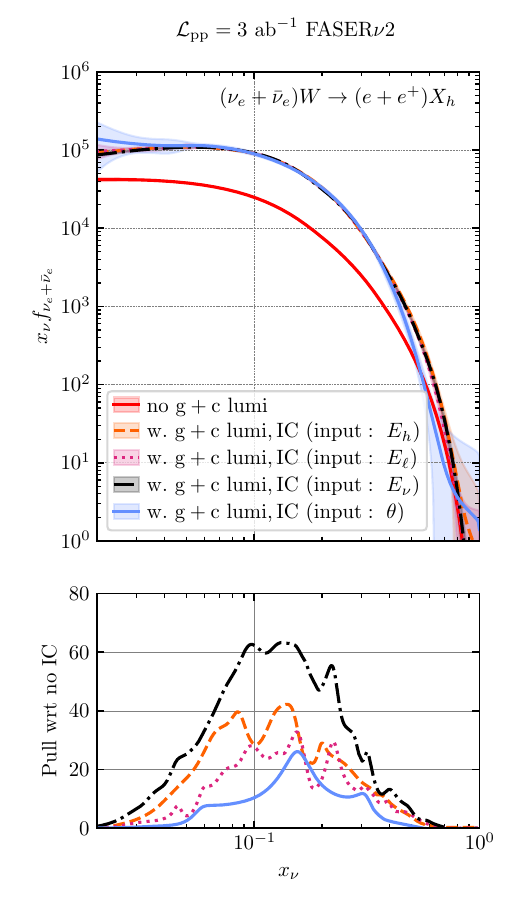}
    \caption{Results of the NN$\nu$flux fits to synthetic data generated from the electron neutrino event yields of Fig.~\ref{fig:IC_events_comp}, without and with the $gQ$ contribution to the charm production cross-sections accounted for.
    We show results for FASER$\nu$ with $\mathcal{L}_{\rm pp}=150$ fb$^{-1}$ (left) and FASER$\nu$2 with $\mathcal{L}_{\rm pp}=3$ ab$^{-1}$ (right panels) for fits based on synthetic data binned in $E_\ell$, $E_h$, $E_\nu$, and $\theta_\ell$ respectively.
    The bottom panels show the statistical pulls Eq.~(\ref{eq:pull_IC_comparison}) evaluated between the results of the various fits as compared to the baseline without the $g+c$ partonic luminosity.
    }
\label{fig:compare_IC_epos}
\end{figure}
%%%%%%%%%%%%%%%%%%%%%%%%%

Fig.~\ref{fig:compare_IC_epos} displays the results of the NN$\nu$flux fits to synthetic data generated from the electron neutrino event yields of Fig.~\ref{fig:IC_events_comp}, without and with the $gQ$ contribution to charm production accounted for.
As mentioned above, the contribution from the $gQ$ initial state is computed using the  CT14 IC BPHS set.
We show results for FASER$\nu$ with $\mathcal{L}_{\rm pp}=150$ fb$^{-1}$ and FASER$\nu$2 with $\mathcal{L}_{\rm pp}=3$ ab$^{-1}$  for fits based on synthetic data binned in $E_\ell$, $E_h$, $E_\nu$, and $\theta_\ell$.
The bottom panels display the pull with respect to the baseline fit, 
\be
\label{eq:pull_IC_comparison}
P_{\rm IC}(x_\nu) = \frac{\Big|\la f^{(\rm gQ,\,IC)}_{\nu_i}(x_\nu)\ra - \la f^{(\rm no\,gQ)}_{\nu_i(x_\nu)}\ra\Big|}{\sqrt{
\sigma^{(\rm gQ,\,IC)2}_{\nu_i}+   \sigma^{(\rm no\,gQ)2}_{\nu_i} 
}} \, ,
\ee
defined as the difference between the fit central values, with the reference being the {\sc\small EPOS+POWHEG} calculation without the $gQ$ contribution, in units of the corresponding fit uncertainties. 

From the results of Fig.~\ref{fig:compare_IC_epos}, one finds that FASER$\nu$ measurements of the electron neutrino event yields based on the Run 3 dataset will already be sensitivity to a possible IC component present in the initial state of the primary proton-proton collision.
An statistical pull of up $P_{\rm IC}\sim 8$ is expected for fits based on the $E_\ell$, $E_h$, and $E_\nu$ distributions for neutrino energies between 700 and 1500 GeV, while this pull quickly decreases for lower and higher energies. 
Therefore, FASER$\nu$ at $\mathcal{L}_{\rm pp}=150$ fb$^{-1}$ has the potential to provide stringent constraints on the intrinsic charm content of the proton via modifications of the electron neutrino forward flux, and eventually could exclude or confirm the perturbative charm scenario.
We recall that we consider only statistical uncertainties in our projections, and hence a detailed estimate of experimental systematic errors is required before drawing any quantitive conclusion from the data.
The right panel of Fig.~\ref{fig:compare_IC_epos} corresponds to the fits to synthetic data for FASER$\nu$2 at the FPF.
The same qualitative trend as for FASER$\nu$ at Run 3 is obtained, though now the pulls are much higher, demonstrating the unique  sensitivity of this experiment to the modelling of the charm production cross-sections at high energies. 
It is also interesting to note that the highest sensitivity is obtained from fitting the $E_\nu$ distribution, and instead fits based on $\theta_\ell$ provide worse constraints, consistently with the closure test analysis of Sect.~\ref{sec:CT}.

The results of Fig.~\ref{fig:compare_IC_epos} demonstrate that the determination of the forward electron neutrino fluxes at FASER and its upgrades may provide conclusive evidence of IC in the proton.
These constraints on IC from neutrino production can be complemented with information provided by muon neutrino~\cite{Cruz-Martinez:2023sdv} and muon scattering~\cite{Francener:2025pnr}, also sensitive to the charm content of the proton, as well as to constraints from other experiments such as the Electron-Ion Collider~\cite{AbdulKhalek:2021gbh,NNPDF:2023tyk}.

\subsection{BSM decays of neutral mesons into neutrinos}
\label{subsec:application_BSM}

In the SM, the branching ratios of neutral hadrons into neutrinos are heavily suppressed, and therefore current experimental bounds on these branching ratios are rather looser than the corresponding SM expectations.
For instance, the branching ratio of neutral pions into electron neutrinos is bounded to be
\be
\label{eq:BR_pi0_SM}
{\rm BR}\lp \pi^0 \to \nu_e + \bar{\nu}_e\rp \le 1.7\times 10^{-6}\qquad {\rm at~the}~95\%~{\rm CL} \, .
\ee
In the SM with massless left-handed neutrinos, this decay is forbidden by angular momentum conservation. 
For massive neutrinos instead, the branching fraction has been estimated to be
\be
{\rm BR}^{(\rm SM)}\lp \pi^0 \to \nu_e + \bar{\nu}_e\rp \simeq 3\cdot 10^{-8} \left(\frac {m_\nu}{m_\pi}\right)^2 \lesssim 3\cdot 10^{-25} ,
\ee
for the case of Dirac neutrinos and a factor of two larger for Majorana neutrinos~\cite{Marciano:1996wy}, where the numerical upper bound is obtained using the current upper limit on the absolute neutrino mass obtained by the KATRIN experiment~\cite{KATRIN:2024cdt} of $m_\nu < 0.45$~eV. 
Similarly considerations apply to the light neutral pseudoscalar mesons $\eta$ and $\eta'$, whose branching factions into electron neutrinos have been constrained to be  
\be
\label{eq:BR_eta_SM}
{\rm BR}\lp \eta \to \nu_e + \bar{\nu}_e\rp \le  10^{-4} 
\qquad \text{and} \qquad
{\rm BR}\lp \eta' \to \nu_e + \bar{\nu}_e\rp \le  6\times 10^{-4}\qquad {\rm at}~95\%~{\rm C.L.} \, .
\ee

The poor experimental constraints on the branching fractions of light neutral hadrons into neutrinos combined with the very large production rates of these same hadrons in the forward region of LHC collisions  implies that  measurements of the electron and muon neutrino fluxes in FASER$\nu$ should impose stringent constraints on BSM scenarios leading to enhanced decay modes of neutral hadrons into neutrinos, potentially improving by a large factor on the current PDG bounds~\cite{ParticleDataGroup:2024cfk}. 

To assess the sensitivity of our NN$\nu$flux approach to discriminate BSM scenarios with enhanced decay rates into neutrinos, we have calculated the neutrino flux with {\sc\small EPOS}+{\sc\small POWHEG} assuming that the branching ratios of $\pi^0$, $\eta$, and $\eta'$ into neutrinos are saturated to their PDG 2024 bounds specified in Eq.~(\ref{eq:BR_pi0_SM}) and Eq.~(\ref{eq:BR_eta_SM}) for electron neutrinos and likewise for the corresponding decay modes into muon and tau neutrinos.
As compared to the corresponding predictions in the SM, enhancing the branching ratios of neutral light mesons into neutrinos has a particularly marked effect for the electron and tau neutrino fluxes, though effects are also expected for muon neutrinos.

\begin{figure}[t]
        \centering
\includegraphics[width=0.87\textwidth]{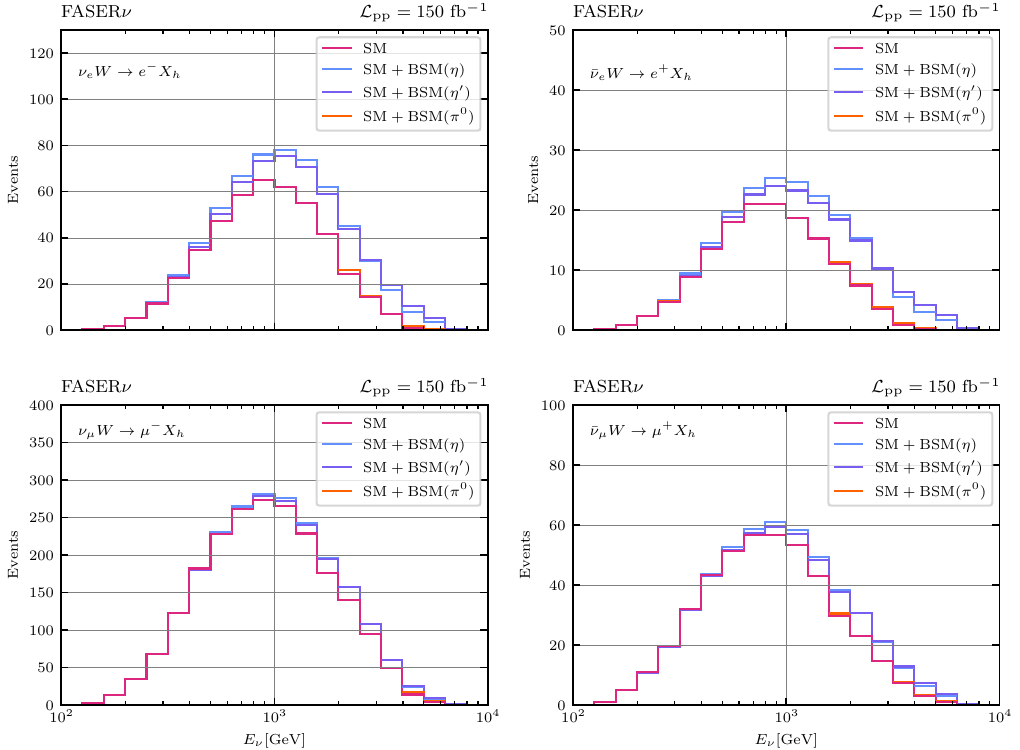}
\caption{The number of interacting electron (top) and muon (bottom) neutrinos (left) and antineutrinos (right panel) in FASER$\nu$ at $\mathcal{L}_{\rm pp}=150$ fb$^{-1}$.
We compare the SM predictions with three BSM scenarios, where the branching ratios of the light neutral mesons $\pi^0$, $\eta$, and $\eta'$ into electron and muon neutrinos are saturated to their PDG  bounds.
 }
\label{fig:neutrino-fluxes-BSM}
\end{figure}

Fig.~\ref{fig:neutrino-fluxes-BSM} displays the number of interacting electron and muon neutrinos and antineutrinos for FASER$\nu$ at $\mathcal{L}_{\rm pp}=150$ fb$^{-1}$ evaluated with {\sc\small EPOS+POWHEG}.
We compare the SM predictions with three BSM scenarios, where the branching ratios of the light neutral mesons $\pi^0$, $\eta$, and $\eta'$ into electron and muon neutrinos are separately saturated to their PDG upper bounds.
From Fig.~\ref{fig:neutrino-fluxes-BSM} one observes that, for electron neutrinos, the contribution from the BSM decay modes of the pseudo-scalar mesons $\eta$ and $\eta'$ grows with the energy, equalling the SM decay channels at $E_\nu\sim 2$ TeV and dominating over them at higher energies.
The contribution of enhanced BSM decays into electron neutrinos of the neutral pion is, on the other hand, almost negligible. 
This result indicates that measurements of the electron neutrino flux at FASER will impose stringent constraints on possible BSM decays of $\eta$ and $\eta'$ mesons into electron neutrinos.
The BSM contributions considered here are instead smaller than the SM prediction for muon neutrinos of any energy, due to the dominance in this channel of neutrinos from kaon decays.

%%%%%%%%%%%%%%%%%%%%%%%%%%%%%%%%
%%%%%%%%%%%%%%%%%%%%%%%%%%%%%%%%%%%%%

\begin{figure}[t]
        \centering
\includegraphics[width=0.41\textwidth]{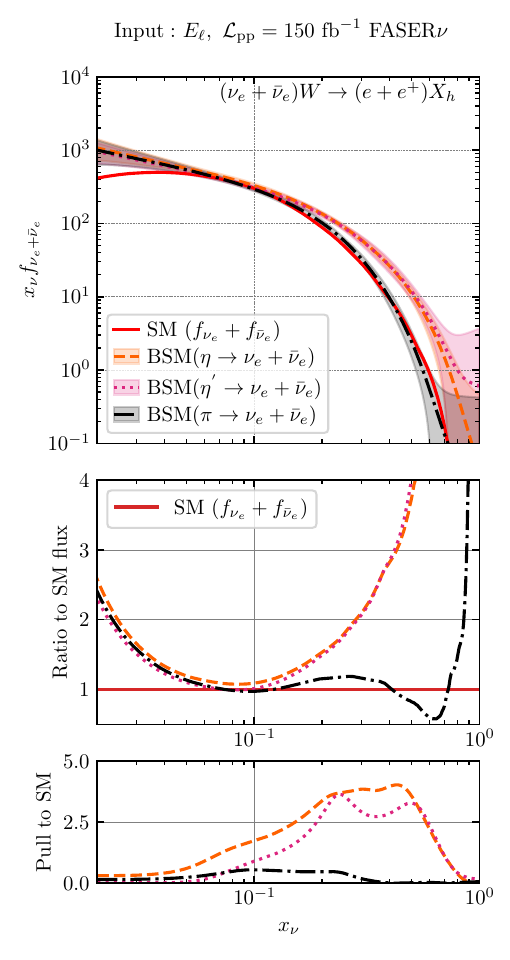}
\includegraphics[width=0.41\textwidth]{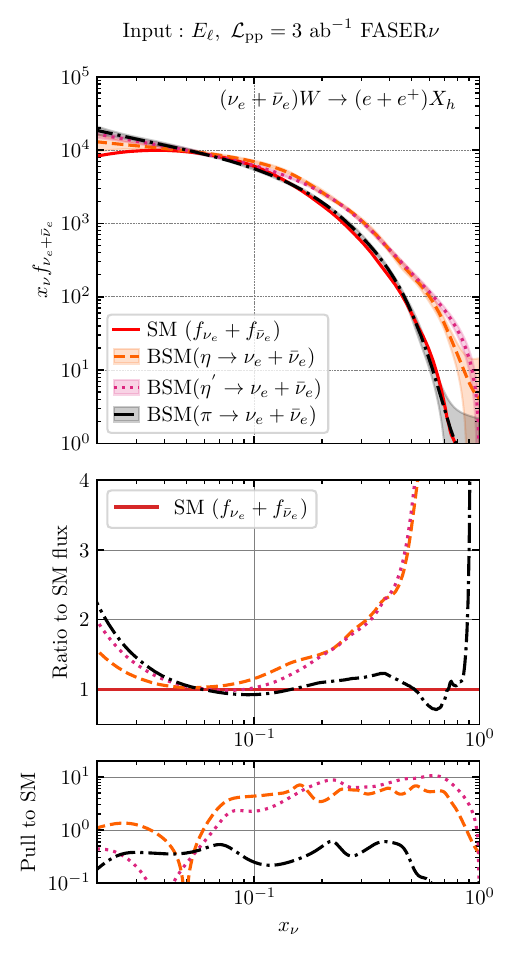}
\caption{Results for the NN$\nu$flux fits to synthetic data for the electron neutrino event rates at FASER$\nu$ with $\mathcal{L}_{\rm pp}=150$ fb$^{-1}$ (left) and   $\mathcal{L}_{\rm pp}=3$ ab$^{-1}$ (right panels).
This synthetic data is generated with {\sc\small EPOS+POWHEG} in the SM and in three BSM scenarios with enhanced branching ratios of the $\pi^0, \eta$ and $\eta'$ mesons into neutrinos consistent with current experimental bounds. 
Results are shown for the sum of neutrino and antineutrino fluxes.
The middle panel shows the fit results normalised to the outcome of the fit based on the SM fluxes, while the bottom panels indicate the pull with respect to the SM defined in Eq.~(\ref{eq:bsm_fluxes}). 
    }
\label{fig:compare_bsm_elec}
\end{figure}

Given that the dominant BSM sensitivity arises from the electron neutrino flux, as shown in Fig.~\ref{fig:neutrino-fluxes-BSM}, we have generated synthetic data for electron neutrino event rates in the SM and with the three BSM scenarios described above. 
Since FASER$\nu$ cannot separate between electron neutrinos and antineutrinos, we present results for the total flux.
Fig.~\ref{fig:compare_bsm_elec} displays the results of the NN$\nu$flux fits to synthetic data generated from the electron neutrino fluxes of Figs.~\ref{fig:neutrino-fluxes-BSM} for FASER$\nu$ with $\mathcal{L}_{\rm pp}=150$ fb$^{-1}$ and $\mathcal{L}_{\rm pp}=3$ ab$^{-1}$.
This synthetic data is generated with {\sc\small EPOS+POWHEG} in the SM and in the three BSM scenarios with enhanced branching ratios of the $\pi^0, \eta$ and $\eta'$ mesons into neutrinos. 
Results are presented for fits carried out to data binned in $E_\ell$, and we have verified that consistent results are obtained when other kinematic variables are used as input. 
The middle panel shows the fit results normalised to the fit based on the SM fluxes, while the bottom panels indicate the pull with respect to the SM,
\be
\label{eq:bsm_fluxes}
P_{\rm BSM}(x_\nu) = \frac{\big|f^{(\rm BSM)}_{\nu_e+\bar{\nu}_e}(x_\nu)-f^{(\rm SM)}_{\nu_e+\bar{\nu}_e}(x_\nu)\big|}{\sqrt{
\sigma^{(\rm BSM)2}_{\nu_e+\bar{\nu}_e}+   \sigma^{(\rm SM)2}_{\nu_e+\bar{\nu}_e} 
}} \, ,
\ee
such that a $P_{\rm BSM}\lsim 1$ indicates that the flux extraction in the BSM and SM scenarios agree with each other within their 68\% CL uncertainties, and higher pulls imply sensitivity the considered BSM scenarios.

%%%%%%%%%%%%%%%%%%%%%%%%%%%%%%%%
%%%%%%%%%%%%%%%%%%%%%%%%%%%%%%%%%%%%%

From Fig.~\ref{fig:compare_bsm_elec} one observes that the pull between the fit results based on the BSM and SM fluxes peaks in the region $1.5~{\rm TeV}\lsim E_\nu\lsim 3.5~{\rm TeV}$, where it reaches an statistical sensitivity of up to $P_{\rm BSM}\sim 4$ for both pseudo-scalar mesons $\eta$ and $\eta'$.
This result confirms that FASER$\nu$ measurements of the electron neutrino fluxes can significantly improve on the current the PDG bounds for the branching ratios of $\eta$ and $\eta'$ into neutrinos.
Likewise, should the considered BSM scenarios be realised in nature, FASER$\nu$ would be able to observe them at the $3\sigma$ (observation) level. 
As in the case of the intrinsic charm analysis of Sect.~\ref{subsec:application_IC}, we emphasize that we consider only statistical uncertainties and that an estimate of detector resolution and other systematic errors is required before the proposed strategy can be fully implemented. 
Due to the higher statistical power, at FASER$\nu$ exposed to the complete HL-LHC luminosity of 3 ab$^{-1}$, the BSM discrimination significance reaches the $10\sigma$ level, hence potentially improving the current PDG bounds by orders of magnitude. 

While here we focus on the $\pi^0, \eta,\eta' \to \nu_\ell+\bar{\nu}_\ell$ decays, we note that similar considerations apply to other possible decay channel into neutrinos.
This includes, for example, decays of the type $\pi/\eta/\eta' \to \nu \nu \gamma $, which could be sizeable in realistic models such as the B$-$L or B$-$3L$_\tau$ gauge groups~\cite{Kling:2020iar, Batell:2021snh}.
Another example is the decay into four neutrinos $\pi^0 \to 2\nu+2\bar\nu$, whose branching fraction has been estimated to be ${\rm BR}^{(\rm SM)}( \pi^0 \to 2\nu_e + 2\bar{\nu}_e) \sim  4\cdot 10^{-23}$ in the SM~\cite{Gao:2018seg} and hence larger than the decay rate into two neutrinos, although still very far from direct experimental bounds.
Furthermore, the possibility to measure rare branching fractions also applies to vector mesons $\rho$, $\omega$ and $\phi$, all of which are profusely produced in the forward region at the LHC~\cite{Kling:2021gos}. 
\section{Summary and outlook}
\label{sec:summary}

In this work we have presented a novel approach for the interpretation of neutrino measurements at the LHC far-forward experiments which realises a data-driven, unbiased extraction of the forward neutrino fluxes by means of machine learning regression.
The main advantage of our pipeline is streamlining the connection between experimental data at the LHC far-forward experiments, on the one hand, and theoretical calculations for forward hadron production in proton-proton collisions, on the other hand. 
Specifically, our results enable the direct comparison of predictions for forward neutrino production with the constraints provided by FASER data while bypassing the need of modelling neutrino scattering, event selection, or detector response.
We have deployed this method to carry out the first direct determination of the LHC muon neutrino flux from the FASER 2024 data and to provide projections for future data-taking periods and for proposed larger detectors. 

To illustrate representative applications of the NN$\nu$flux method we have presented three case studies. 
First, to benchmark predictions of event generators of forward particle production, demonstrating that current and future FASER measurements exhibit decisive discrimination sensitivity.
Second, to scrutinize the charm content of the proton, by means of the characteristic  enhancement of the $gQ$ partonic luminosity in $D$-meson production arising in the presence of intrinsic charm.
Third, to provide stringent constraints on BSM scenarios leading to enhanced branching ratios for the decays of neutral mesons into neutrinos.
Several other applications could be envisaged for future work, such as discriminating between DGLAP-based and BFKL-based or non-linear QCD calculations of forward $D$-meson production, constraining models of $D$-meson fragmentation, and providing dedicated tunes of forward MC generators for cosmic ray experiments. 

From the methodology viewpoint, it would be possible to extend our flux extraction technique first to the  FASER rapidity measurements, to check consistency with the $E_\nu$-based determination of Sect.~\ref{sec:results_faser}, and subsequently to a two-dimensional determination which accounts for both the energy and rapidity dependence of the fluxes, following App.~\ref{app:2d-lumis}.
Furthermore, given that our pipeline provides NLO-accurate predictions for exclusive final states, one may attempt flux extractions based on detector-level observables in the fiducial region, bypassing the need of unfolding to the initial--state kinematics. 
Finally, once a sufficiently large data sample becomes available, one would like to combine the methodology presented in this work with the the PDF determination studies of~Ref.\cite{Cruz-Martinez:2023sdv} to realise a simultaneous extraction of the neutrino flux and of the PDFs, benefitting that different observables, detectors, and kinematic variables provide complementary constraints on the two analysis targets. 

Another future goal of the NN$\nu$flux methodology is bridging the gap between  neutrino measurements at FASER and predictions for the prompt neutrino flux from charm decays at neutrino telescopes such as IceCube and KM3NET. 
To achieve this, once a energy- and rapidity-dependent neutrino PDF  is determined for both electron and tau neutrinos, one would use theory simulations of $D$-meson fragmentation and decay to reconstruct the original charm production cross-section $d^2\sigma_{c\bar{c}}/dy_cdp_T^c$ and apply it, convoluted with the cosmic ray fluxes, to predict the prompt flux in the region relevant for cosmic neutrinos.
This data-driven approach has as main benefit bypassing the large uncertainties, arising both from PDFs and higher-order QCD corrections, which affect a first-principle calculation of charm production in the forward region.
Such application would be pivotal to inform the data analysis of ongoing and future astroparticle physics experiments involving ultra-high-energy neutrinos. 

As the LHC far-forward experiments (and their Run 4 upgrades) continue to deliver results, developing novel analysis techniques to streamline their physics interpretation becomes a priority.
Here we have contributed to this program by lowering the threshold to connect experimental data with theory predictions of forward hadron and neutrino production at high energies, and therefore facilitate novel SM and BSM interpretations of these measurements.   

\vspace{0.2cm}
\begin{center}
\rule{0.6\linewidth}{0.4pt}
\end{center}
\vspace{0.2cm}

\noindent
The results of this work can be reproduced and extended by means of the open-source NN$\nu$flux code available:
\begin{center}
\url{https://github.com/LHCfitNikhef/nnfluxnu}
\end{center}
which includes run cards and example analysis files illustrating its main functionalities. 
Instructions on how to install and use the code are available from:
\begin{center}
\url{https://jukkajohn.github.io/documentation}
\end{center}

\noindent
The calculation of the neutrino scattering event yields used in this work are based on the {\sc\small POWHEG~DIS} code from~\cite{vanBeekveld:2024ziz} extended to generate fast interpolation FK-tables for arbitrary final-state variables and fiducial selection,  available from
\begin{center}
\url{https://github.com/LHCfitNikhef/powheg-dis/tree/fktable}
\end{center}

\subsection*{Acknowledgements} 
We thank Rafal Maciula and Antoni Szczurek for sharing the files used to make predictions for $D$-meson production in the presence of intrinsic charm. 
The work of J.~R. is partially supported by the Dutch Science Council (NWO).
The work of F.~K. is supported in part by Heising-Simons Foundation Grant 2020-1840 and in part by U.S. National Science Foundation Grant PHY-2210283.

\appendix
\section{Neutrino PDFs with $E_\nu$ and $y_\nu$ dependence}
\label{app:2d-lumis}

In Sect.~\ref{sec:formalism} we have described the NN$\nu$flux methodology in the case of the $E_\nu$-dependent fluxes. 
Its extension to double-differential fits in which we retain both the neutrino energy and rapidity dependence of the fluxes is technically straightforward. 
We define a two-dimensional analogue of the neutrino PDF of Eq.~(\ref{eq:neutrino_pdf_definition}) as
\be
\label{eq:neutrino_pdf_definition_2D}
\widetilde{f}_{\nu_i}(x_{\nu},y_\nu)\equiv \frac{\sqrt{s_{\rm pp}}}{2} \frac{d^2N_{\nu_i}(E_\nu,y_\nu)}{dE_{\nu}dy_\nu} \, ,\qquad i = e,\mu,\tau \, ,
\ee
which encapsulates the dependence of the fluxes on the neutrino rapidity (that is, on its non-zero transverse momentum with respect to the LoS). 
We can now express the neutrino PDF in terms of a basis of 2D interpolating functions, namely
\begin{equation}
\label{eq:interpolated_expression_fluxes_v2}
    \widetilde{f}_{\nu_i}(x_\nu,y_\nu) \simeq  \sum_{\alpha=1}^{n_{x}}
\sum_{\beta=1}^{n_{y}}\widetilde{f}_{\nu_i}(x_{\nu,\alpha},y_{\nu,\beta})I_\alpha(x_\nu)I_\beta({y_{\nu}}) \, ,
\end{equation}
and complement the previous $x_\nu$ grid with a second grid with $n_y$ nodes $\{ y_{\nu,\beta}\}$ for the dependence on the neutrino rapidity. 
Let's then take $n_{E_\nu}$ and $m_{y_\nu}$
bins in the neutrino energy and rapidity,
\be
E^{\rm (min)}_{\nu,j} \le E_{\nu,j}
\le E^{\rm (max)}_{\nu,j} \, ,\qquad
y^{\rm (min)}_{\nu,k} \le y_{\nu,k}
\le y^{\rm (max)}_{\nu,k}\, ,
\ee
so the double-differential event rates are given by
\bea
\label{eq:event_yields_calculation_v13}
 N_{\rm int}^{(\nu_i)}(E_{\nu,j}, y_{\nu,k})&=&
  \int_{E_{\nu,j}^{\rm (min)}}^{E_{\nu,j}^{\rm (max)}}
 dE_{\nu}
 \int_{y_{\nu,k}^{\rm (min)}}^{y_{\nu,k}^{\rm (max)}}
 dy_\nu\,
 \int_{Q^{2}_{0}}^{2m_NE_\nu}
  dQ^2
 \int_{Q^2/2m_NE_\nu}^{1}
  dx \\
  &\times& \nonumber \lp 
   \frac{2n_T L_T}{\sqrt{s_{\rm pp}}} \widetilde{f}_{\nu_i}(x_{\nu},y_\nu)  \frac{d^2\sigma_{\rm NLO+PS}^{\nu_i A}(x,Q^2,E_{\nu})}{dxdQ^2}\Bigg|_{\rm fid}\rp  \\
  & = &
  \sum_{\alpha=1}^{n_{x}}
\sum_{\beta=1}^{n_{y}}\widetilde{f}_{\nu_i}(x_{\nu,\alpha},y_{\nu,\beta})
  \int_{E_{\nu,j}^{\rm (min)}}^{E_{\nu,j}^{\rm (max)}}
 dE_{\nu}
 \int_{y_{\nu,k}^{\rm (min)}}^{y_{\nu,k}^{\rm (max)}}
 dy_\nu\,
 \int_{Q^{2}_{0}}^{2m_NE_\nu}
  dQ^2
 \int_{Q^2/2m_NE_\nu}^{1}
  dx \nonumber\\
 &\times&  \lp  \frac{2n_T L_T}{\sqrt{s_{\rm pp}}} 
I_\alpha(x_\nu)I_\beta(y_{\nu}) \frac{d^2\sigma_{\rm NLO+PS}^{\nu_i A}(x,Q^2,E_{\nu})}{dxdQ^2}\Bigg|_{\rm fid}\rp  \nonumber\\
  &\equiv & \sum_{\alpha=1}^{n_{x}}
\sum_{\beta=1}^{n_{y}}\widetilde{f}_{\nu_i}(x_{\nu,\alpha},y_{\nu,\beta}){\rm FK}_{\alpha,\beta,j,k}\, ,
    \nonumber
\eea
where now the FK-table is a tensor with four components: two labelling the bin in $E_\nu$
and $y_\nu$, the other two labelling the
interpolation basis polynomial in the $E_\nu$
and $y_\nu$ directions, respectively:
\be
\nonumber
{\rm FK}_{\alpha,\beta,j,k} = \frac{2n_TL_T}{\sqrt{s_{\rm pp}}} \int_{E_{\nu,j}^{\rm (min)}}^{E_{\nu,j}^{\rm (max)}}
 dE_{\nu}
 \int_{y_{\nu,k}^{\rm (min)}}^{y_{\nu,k}^{\rm (max)}}
 dy_\nu\,
 \int_{Q^{2}_{0}}^{2m_NE_\nu}
  dQ^2
 \int_{x_0(Q^2)}^{1}
  dx  \,
I_\alpha(x_\nu)I_\beta(y_{\nu}) \frac{d^2\sigma_{\rm NLO+PS}^{\nu_i A}(x,Q^2,E_{\nu})}{dxdQ^2}\Bigg|_{\rm fid} \, .
\ee
Therefore one ends up with an expression for the binned event rates of the form
\be
N_{\rm int}^{(\nu_i)}(E_{\nu,j}, y_{\nu,k}) = \sum_{\alpha=1}^{n_{x}}
\sum_{\beta=1}^{n_{y}}\widetilde{f}_{\nu_i}(x_{\nu,\alpha},y_{\nu,\beta}){\rm FK}_{\alpha,\beta,j,k} \, ,
\ee
which can be used to fit the rapidity-dependent neutrino PDF by parametrizing it as
\be
\label{eq:neutrino_pdf_definition_outlook}
\widetilde{f}_{\nu_i}(x_{\nu},y_{\nu})= \frac{\sqrt{s_{\rm pp}}}{2} \frac{d^2N_{\nu_i}(E_\nu)}{dE_{\nu}dy_\nu} =  x_{\nu}^{a_i} (1-x_\nu)^{b_i}{\rm NN}_{\nu_i}(x_\nu,y_\nu) \, ,
\ee
with a two-input neural network as in the case of the nNNPDF fits~\cite{AbdulKhalek:2022fyi} or the NNSF$\nu$ neutrino structure function determination~\cite{Candido:2023utz}.
\bibliographystyle{utphys}
\bibliography{ForwardNuFluxes}

\end{document}